\documentclass[usegraphicx,useAMS,usenatbib]{mn2e}
\usepackage{url}
\title[Ram pressure stripping and magnetised turbulent ISM]
{Ram pressure stripping in elliptical galaxies: II. magnetic field effects}
\author[M.-S. Shin, \& M. Ruszkowski]
  {Min-Su~Shin,$^{1}$\thanks{E-mail: 
  Min-Su.Shin@astro.ox.ac.uk, 
  mateuszr@umich.edu}
  Mateusz~Ruszkowski$^{2}$\\
  $^1$Sub-department of Astrophysics, Department of Physics, University of Oxford, Keble Road, Oxford OX1 3RH\\
  $^2$Department of Astronomy, The University of Michigan, 500 Church Street, Ann Arbor, MI 48109, USA\\
  }
\date{Released 2009 Xxxxx XX}

\begin{document}

\date{Accepted ... Received ..; in original form ..}

\maketitle

\begin{abstract}
We investigate the effects of magnetic fields and turbulence on ram 
pressure stripping in elliptical galaxies using ideal 
magnetohydrodynamics simulations.
We consider weakly-magnetised interstellar medium (ISM) 
characterised by subsonic turbulence, 
and two orientations of the magnetic fields in the
intracluster medium (ICM) -- parallel and perpendicular to the
direction of the galaxy motion through the ICM. 
While the stronger turbulence enhances the ram 
pressure stripping mass loss, the magnetic fields tend to suppress the
stripping rates, and the suppression is stronger for parallel fields. 
However, the effect of magnetic fields on the mass
stripping rate is mild. Nevertheless, the morphology of the stripping 
tails depends significantly on the direction of 
the ICM magnetic field. 
The effect of the magnetic field geometry
on the tail morphology is much stronger than that of the level of the ISM
turbulence. The tail has a highly collimated shape for parallel fields, 
while it has a sheet-like morphology in the plane of the ICM magnetic field 
for perpendicular fields.
The magnetic field in the tail is amplified 
irrespectively of the orientation of
the ICM field.
More strongly magnetised regions in the ram
pressure stripping tails are expected to 
have systematically higher metallicity 
due to the strong concentration of the stripped ISM
than the less magnetised regions. 
Strong dependence of the morphology of the stripped
ISM on the magnetic field could
potentially be used to constrain the relative orientation of the
ram pressure direction and the dominant component of the ICM magnetic field.
\end{abstract}


\section{INTRODUCTION}

A number of environmental processes can alter galaxy evolution in
cluster or group environments \citep{2004IAUS..217..440,2006PASP..118..517B}. 
Among these, ram pressure stripping process plays a particularly
important role in galaxy evolution. This process strongly depends on
galactic environment, and is especially strong when 
galaxies interact with high density ICM and move at large 
relative velocity with respect to the ICM.
The impact of this process was first 
demonstrated comprehensively in a classic paper 
by \citet{1972ApJ...176....1G}.
Since then, a number of studies considered consequences of ram
pressure stripping for galaxy
evolution using numerical models.
For example, these simulations considered a number of important 
issues such as:
replenishment of the ISM lost 
from elliptical galaxies due to ram pressure stripping
\citep{1987ApJ...316..530G}, possibility of re-accretion of the stripped ISM 
\citep{1994ApJ...437...83B}, complex time dependence of the ISM loss 
on the galaxy cluster environment \citep{1999MNRAS.310..663S}, 
inadequacy of analytical estimation for predicting 
the true ISM stripping rates \citep[e.g.][]{2007MNRAS.380.1399R}, 
the effect of realistic ICM substructures on the ISM mass loss in 
cosmological setups \citep{2008ApJ...684L...9T}, 
impact of the orientation of disc galaxies 
with respect to their orbital trajectory \citep{2009A&A...500..693J}, and 
star formation in the context of ram pressure stripping  
\citep{2014MNRAS.443L.114R}.

As pointed out in \citet[][hereafter, Paper I]{paper1}, 
most simulations of ram 
pressure stripping did not include the effect of 
non-thermal energy components in the ISM and ICM. 
In the ISM, the non-thermal energy contributions 
include turbulent kinetic energy, 
magnetic fields, and cosmic-rays. In Paper I, we investigated
the effect of the ISM turbulence on ram pressure 
stripping in elliptical galaxies, showing that 
the ISM mass loss is enhanced by including the ISM turbulence. 
\citet{2013MNRAS.436.3021C} 
support our results found in Paper I, using sub-grid turbulence 
models.
However, our previous simulations 
ignored ISM and ICM magnetic fields, which are known to be present 
\citep{1990A&A...233..417L,1996MNRAS.279..229M,
2002RvMP74.775W,2011Vallee,2012ApJ...757..123A}.

The process of ram pressure stripping in clusters or groups is to some
extent analogous to the interaction of the heliosphere with
the local magnetised ISM through which the Sun is moving. The ram pressure of
the solar wind competes with that corresponding to the
circum-heliospheric ISM, and prevents the ISM from completely blowing
away the solar material \citep[see][for a review]{2011ARA&A..49..237F}. 
The geometry of the magnetic field in the local ISM and heliosphere, as 
well as the properties of turbulence, are not well understood. 
Nevertheless, the strength and geometry of the magnetic fields are
thought to be important factors in regulating the mass exchange between 
the local ISM and heliosphere \citep{2009Natur.462.1036O,
2011ApJ...734...71O,2011ApJ...742..104P}.

The analogy between galactic ram pressure stripping 
and the interaction of the heliosphere 
with the local ISM suggests that the magnetic fields in the ISM and
ICM should also affect the efficiency of the ram pressure 
stripping and the spatial distributions of the gas and magnetic fields. 
\citet{2014ApJ...784...75R} 
simulated ram pressure stripping in the case of a disk galaxy 
interacting with weakly magnetised ICM. They considered different
orientations of the ICM magnetic fields and relative orientations
of the disk with respect to the ram pressure direction.
They find that the presence of magnetic fields has a strong effect on
the tail morphology -- it leads to 
filamentary tails rather than clumpy ones 
predicted by purely hydrodynamic simulations. 
In particular, their results show that the presence of magnetic fields
may lead to the double tails similar to the ones seen in ESO 137-001
and ESO 137-002 \citep{2006ApJ...637L..81S,2007ApJ...671..190S,
2013ApJ...777..122Z}.
\citet{2003A&A...402..879O} also studied the effects of magnetic fields 
on ram pressure stripping of 
late-type galaxies. However, their simulations did not 
include dynamical coupling of magnetic 
fields to the ISM and neglected ICM fields. 

Since the magnetic field strength and its spatial distribution in
elliptical galaxies are quite different from those 
representative of late-type 
galaxies, and because the distribution of the gas in late-type galaxies 
(relatively dense and flat gaseous disk and tenuous hot halo gas) 
is unlike what is seen in elliptical galaxies, 
the simulations of ram pressure stripping in disk galaxies 
cannot be used to understand the impact of the ram pressure 
stripping on elliptical galaxies.

In this second paper in the series, we aim to explore how the
weakly-magnetised turbulent ISM and the magnetic field in the
ICM affect the ram pressure stripping process in elliptical galaxies. 
We systematically investigate trends in mass stripping 
rates and morphology of the ram pressure stripping tails 
with the magnetic field strength, geometry, and ISM turbulence strength.
While late-type galaxies have well-ordered large-scale structures 
in the spatial distribution of 
magnetic fields, early-type galaxies are expected 
to have highly tangled fields 
\citep[see][for a review]{2011AIPC.1381..117B}. 
In elliptical galaxies, the ISM pressure support against gravity is
thought to come partially from subsonic turbulent motions 
\citep{2009MNRAS.398...23W,
2010MNRAS.406..354O,2011MNRAS.410.1797S,
2012A&A...539A..34D,2013MNRAS.430.1516H}. 
Elliptical galaxies are also commonly found 
in the inner regions of galaxy clusters, where the ICM is 
relatively dense. This hot ICM 
is thought to be weakly magnetised and turbulent
\citep{1994AJ....107.1942T,1994AJ....108.1523G,
2004IJMPD..13.1549G,2005MNRAS.358..139F,
2006MNRAS.366.1437S,2012A&A...540A..38V}. 
The topology of the magnetic fields in the ICM is not known but some
studies indicate that the fields can be coherent on the scales
comparable to or larger than the size of an elliptical galaxy 
\citep[e.g.,][]{2011A&A...529A..13K,2012A&A.540A.38V}.
Therefore, it is necessary to incorporate  the
effects of the magnetised ICM and ISM
in ram pressure stripping simulations of elliptical galaxies.

For the sake of simplicity, in this exploratory study we consider only 
very simplified geometry of the uniform ICM fields -- parallel and 
perpendicular to the direction of the ram pressure. 
The objective of this investigation is to assess how sensitive the 
ram pressure stripping is to the field geometry in 
addition to the strength of the ISM turbulence. 
We focus on the effect of magnetic fields 
on the mass loss from galaxies  
and on the observable features of galaxies and their stripping tails. 
In future work, we will consider realistic distributions of the turbulent ICM
magnetic fields, and the results presented here will help us to interpret 
future simulations. 
In order to highlight the differences between purely hydrodynamic 
simulations and MHD simulations, we compare main results presented
here to those from our Paper I.

Our new simulations are complementary 
to the studies of cold fronts in clusters 
\citep[e.g.,][]{Lyutikov2006,2007ApJ...663..816A,
2007MNRAS.379.1325X,2008ApJ...677..993D, Takizawa2005}. 
The physical scale and situation envisaged in our simulations are 
different. Ram pressure leads to the removal of the ISM 
from the galaxies, and the ICM is allowed to penetrate the galactic
regions initially occupied by the ISM.
In our simulations, the distributions of magnetic fields, temperature, 
and density are smooth on the boundary between the ISM and ICM, 
while most studies of the moving substructure in clusters assume 
a contact discontinuity or a sharp ISM/ICM boundary. 

This paper is organised as follows. In Section 2, 
we discuss the simulation setup. Results 
are presented in Section 3. Finally, we discuss 
the implications of the results 
and conclude in Section 4. In the Appendix, 
we present the results of additional simulations to 
validate our main simulation.

\section{SIMULATIONS}

We use {\tt FLASH4-alpha} code to solve ideal MHD equations 
\citep{2000ApJS..131..273F,2009JCoPh.228..952L}.
We adopt unsplit MHD solver and use Roe Riemann solver with 
van Leer flux limiter. 
The initial distribution of the ISM is assumed to be spherical and in
a hydrostatic equilibrium with static gravitational field of the
galaxy. The stellar component of the gravitational field is described by 
a spherical Jaffe model. The dark matter distribution is chosen such that the 
total radial mass profile scales with radius $r$ as $r^{-2}$ 
\citep[e.g.,][]{2009MNRAS.393..491C,2009ApJ...699...89C,2010ApJ...711..268S}. 
We truncate the gravitational field 
beyond a truncation radius $R_{t}=100$ kpc. 
This truncation approximates the effect of cluster tidal field on 
the galactic gravitational field 
\citep[e.g.,][]{2009ApJ...696.1771L}. The total stellar mass of the galaxy 
is about ${\rm 10^{11} M_{\odot}}$
and the mass of the dark matter halo is equal to the stellar mass inside
the stellar-mass effective radius 
3 kpc \citep{2012ApJ...748....2D}. The centre of the galaxy is located 
at (0, 0, 0) and the simulation box covers a wide range 
from $-320$ to $+830$ kpc along $x$-axis, 
and from $-256$ to $256$ kpc along other axes.

The density and temperature distributions of the ISM and ICM are
continuous at $R_{t}$. The ICM density and temperature 
are constant and equal to 
$3 \times 10^{-28}$ ${\rm g ~ cm^{-3}}$ and $2 \times 10^{7}$ K, 
respectively. The initial temperature distribution of the ISM is
\begin{equation}
T(r) = \left\{ \begin{array}{rl}
T_{i}                             &  \mbox{ if $r < r_{i}$} \\
2T_{0}/(1 + (r / r_{0})^{-3})   &  \mbox{ otherwise}, 
\end{array} \right.
\end{equation}
where $T_{i}=8 \times 10^{6}$ K, $r_{i}=50.9$ kpc, 
$T_{0}=1.3\times 10^{7}$ K, and $r_{0}=66.6$ kpc. For the
sake of simplicity, we assume the ideal gas equation of state 
for completely ionised gas with solar metallicity in both ISM and ICM. 

\begin{table*}
\begin{minipage}{130mm}
\caption{Simulation runs.}
\label{tab:run}
\begin{tabular}{lc||cc|cc}
\hline
& & \multicolumn{2}{c}{Case PA} & \multicolumn{2}{c}{Case PP} \\
Name & ISM injection energy & ISM 1D RMS velocity & $ <\beta> $ & ISM 1D RMS velocity & $ <\beta> $ \\
& (${\rm cm^{2} ~ s^{-3}}$) & (Mach number) & & (Mach number) & \\
\hline
Run 0 & $2.5 \times 10^{-8}$ & 0.022 & 44.8 & 0.023 & 24.0 \\
Run 1 & $2.5 \times 10^{-7}$ & 0.028 & 53.2 & 0.027 & 28.2 \\
Run 2 & $5.0 \times 10^{-7}$ & 0.036 & 45.4 & 0.033 & 25.8 \\
Run 3 & $1.0 \times 10^{-6}$ & 0.047 & 35.3 & 0.042 & 25.4 \\
Run 4 & $2.0 \times 10^{-6}$ & 0.060 & 30.8 & 0.055 & 28.0 \\
Run 5 & $4.0 \times 10^{-6}$ & 0.084 & 34.5 & 0.078 & 33.0 \\
\hline
\end{tabular}

\medskip
{ISM injection energy} represents energy per stirring mode. ISM 1D RMS
velocity is measured along the $z$-axis and is 
mass-weighted. $<\beta>$ is a mass-weighted plasma beta parameter averaged inside $R_{t}$.
\end{minipage}
\end{table*}

Initially, both ISM and ICM include a weak magnetic field 
of $1.44 \times 10^{-6}$ Gauss, 
which corresponds to plasma beta ($\beta ~=~ 
(n k_{B} T) / (B^{2} / (8 \pi))$) 
of about 650 and 10 
for the ISM at the galactic centre and ICM, respectively. 
Because $\beta\gg 1$, the initial density and temperature
profiles do not depend on the strength of these initial fields.
Since we also drive subsonic turbulence in the ISM, any small
departures from perfect hydrostatic equilibrium due to magnetic forces
are negligible compared to the departures from
hydrostatic equilibrium due to driven subsonic turbulence in the ISM.

We consider two different directions of the magnetic field. In one set 
of simulations, the initial magnetic field is oriented along the $x$-axis, 
which is also the direction of the ram pressure. We call 
this set of simulations Case PA. The other set of simulations 
(Case PP) assumes the initial magnetic field along the $y$-axis. 

As in Paper I, we use a modified stirring module in the {\tt FLASH} code 
\citep{1988CF.....16..257E,DBLP:journals/ibmrd/FisherKLDPCCCFPAARGASRGN08,
2010ApJ...713.1332R}. In order to generate
weakly magnetised turbulent ISM, we perturb the initially 
hydrostatic ISM by adding kinetic energy to the gas at 
six different injection rates (see Table \ref{tab:run}). 
The injection rates are per stirring
mode. We use 152 modes corresponding to wavelengths ranging 
from 49 to 50 kpc. This
stirring occurs only for $r<R_{t}$, and continues 
for the entire duration of the simulations. 
At 0.5 Gyr, the ICM of the constant density, temperature, and magnetic field 
starts to flow into the simulation domain from the -$x$ boundary and
begins to exert ram pressure on the galaxy. 
At this time, the turbulence in the ISM has reached a steady state
characterised by plasma $\beta$ and velocity dispersions summarised in 
Table \ref{tab:run}. At 0.5 Gyr, the ISM magnetic field in Case PA is
weaker than in Case PP despite the fact that the
same amount of kinetic energy is injected in both cases 
over the same amount of time. This happens because, during this
initial stage, the boundary conditions are fixed in
time are continuously reset at the -$x$ boundary. 
In Case PA, the magnetic field lines at the boundary 
are connected to the rest of the 
volume (at the boundary the fields are perpendicular to it) and,
the fields are consequently not amplified as efficiently as in Case
PP where the field lines do not intersect the $-x$ boundary.
The magnetic field pressure is negligible
compared to gas pressure in all runs initially. 
The simulations are evolved for 6 Gyr, including the initial 0.5 Gyr
spent on stirring the ISM before the onset of the ICM inflow.

When investigating the effects of magnetic fields on ram pressure 
stripping, we compare Runs 0 and 2 of 
Case A in Paper I to Runs 0 and 3 of Cases PA and PP in this paper.
Because the strength of turbulence in the ISM, which is measured 
by 1D RMS velocity dispersions, is very similar in these
runs (see Table 1 in Paper I and 
Table \ref{tab:run} here), they are useful for isolating  
the effects of turbulence in the ISM.

Unless stated otherwise, we set the inflow 
velocity of the ICM at $\sim 170$ ${\rm km / s}$, 
which is equal to Mach number of 0.25 with respect to the 
sound speed of the ICM. 
For Runs 0 and 5, we also investigate the effect of 
stronger ram pressure by increasing the speed of the ICM three times 
(i.e., by increasing the strength of the ram pressure nine times). 
These additional runs are denoted as Run 0h and Run 5h. 
Subsonic speeds may be typical 
of elliptical galaxies after they have been  
completely incorporated into galaxy clusters 
\citep[e.g.,][]{1998A&A...331..439A,2008ApJ...676..218H}. 
We note that this may not be the case  
for both late-type galaxies and galaxies in the process of falling into
clusters for the first time 
\citep[e.g.,][]{2005MNRAS.358..139F,2005ApJ...621..663M,2006ApJ...644..155M}.

We use two different kinds of tracers in the simulations -- tracer
particles and a passive scalar -- in order
to examine the mixing between the ISM and ICM, and to understand 
how the ISM is transported from the galaxy to the stripping tail. 
Passively moving particles are included with separate tags for the 
ISM and ICM. Including such particles allows us to track the origin of
the particles and gas in the tail \citep{2009AnRFM..41..375T,
2007MNRAS.374..787H}. We distribute 8,168 ISM particles 
uniformly inside $R_{t}$. We note that as the particles are dispersed
through the volume, it is increasingly difficult to densely sample the
growing tail volume with a finite number of
particles. Therefore, we also introduce a passively advected scalar
which we call ``colour''. We use this quantity 
to estimate what fraction of mass in simulation cells comes from the ISM. 
If a cell contains only the ISM, 
the value of colour is, by definition, equal to 1. 
Colour does not allow us to track gas history. However, the distribution of 
colour shows how well two different kinds 
of gases are mixed \citep[e.g.,][]{2008ApJ...680..336S}. 

We use the colour as a refinement variable. 
Because there is no absolute rule for the best 
variables of refinement and refinement 
conditions 
\citep[][for discussion]{1989JCoPh..82...64B, 2010JCoAM.233.3139L},
we refine cells which have 
strong spatial variations of the colour above {\tt refine\_cutoff}=0.8, 
and derefine cells which have weak 
variations below {\tt derefine\_cutoff}=0.2, 
and use  {\tt refine\_filter}$=10^{-2}$ in {\tt FLASH4-alpha} 
\footnote{See 
\url{http://www.asci.uchicago.edu/site/flashcode/user_support/flash3_ug_3p3/} 
for definitions of these simulation parameters.}. 
We use adaptive mesh refinement only for $r>R_{t}$. The refinement level is 
allowed to vary between 3 and 6 levels and the maximum resolution 
is 1 kpc. For $r<R_{t}$ we fix the spatial resolution at 2 kpc. 
We note that this refinement rule may affect structures found 
at $r>R_{t}$ quantitatively since it changes spatial resolution of
the stripped ISM and the smallest scale of turbulent structures 
and magnetic-field amplification in that region. 
However, this configuration is the same as in our Paper I 
(i.e., pure hydro runs), allowing us to perform meaningful 
{\it relative} comparisons.

\begin{figure*}
\includegraphics[width=165mm]{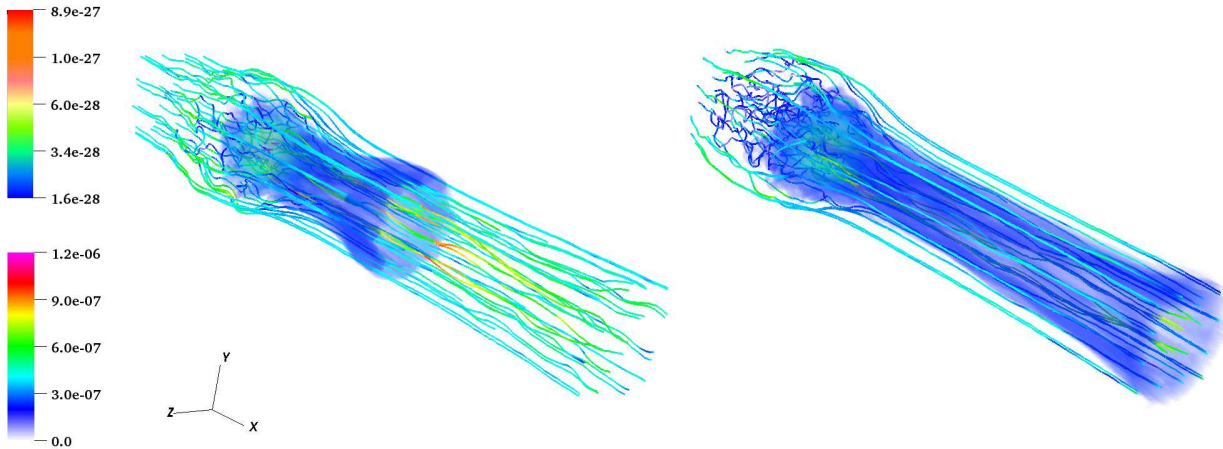}
\caption{Gas density distributions for the ISM mass fractions 
above 1\% and magnetic field magnitude in Run 5, Case PA 
at 3 ({\it left}) and 6 ({\it right}) Gyr. The top colour bar 
shows density in units of  ${\rm g ~ cm^{-3}}$.
The bottom colour bar shows magnetic field in $\sqrt{4 \pi}$ Gauss. 
Therefore, 1 in this unit corresponds to $\sqrt{4 \pi}$ Gauss.
The magnetic field lines extend from about -320 to 650 kpc along $x$-axis. 
}
\label{fig:3D_para}
\end{figure*}

\begin{figure*}
\includegraphics[width=165mm]{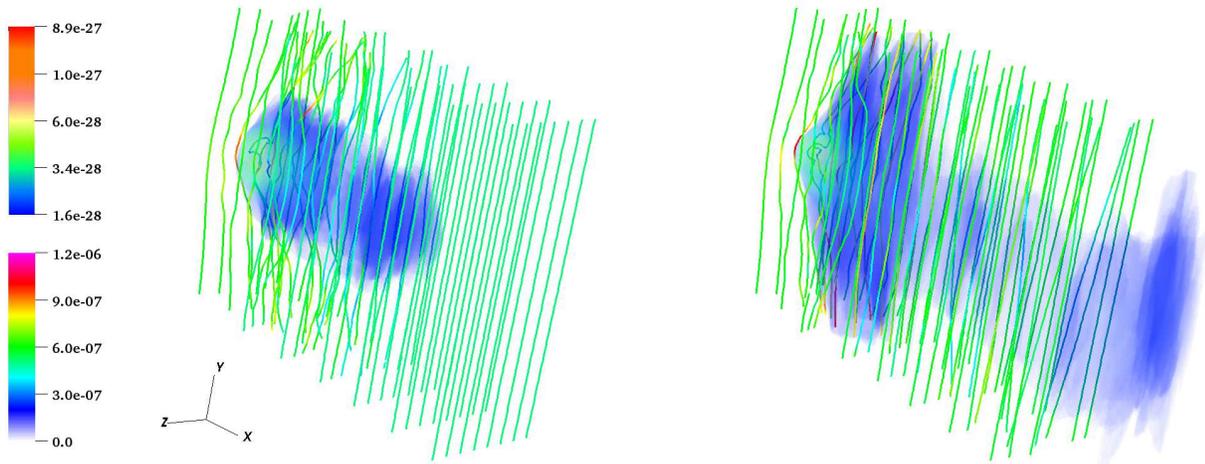}
\caption{Same as Figure \ref{fig:3D_para} but for Case PP. 
Significant magnetic field is found near the front side of the galaxy
exposed to the inflowing ICM. The field is also amplified as it wraps around
the galaxy. The direction of the tail is largely co-aligned 
with the magnetic field lines in the tail. 
The spatial scale is the same as in Figure 
\ref{fig:3D_para}. The field lines are plotted between -190 and 190
kpc along the $y$-axis.
}
\label{fig:3D_perp}
\end{figure*}

\section{RESULTS}

We performed twelve simulations for the low-speed ICM velocity $\sim
170$ ${\rm km / s}$ and four runs for three times higher speeds.
These simulations consider different ICM field orientations and
turbulence strengths. We focus on the relative comparisons 
between these runs, but also examine 
the differences between purely hydrodynamical simulations 
presented in Paper I and the MHD simulations presented here.

\subsection{Overall evolution}

Figures \ref{fig:3D_para} and \ref{fig:3D_perp} demonstrate that there
are dramatic morphological differences in the gas distributions 
between Case PA and PP. Case PA simulation reveals 
a long tail stretched in the direction of the ram pressure, 
while Case PP produces a tail flattened in the $x-y$ plane, i.e., in the
plane of the incoming ICM magnetic field.
Magnetic fields are strongly amplified along the x-axis behind the galaxy 
in Case PA. On the other hand, in Case PP, the magnetic fields are amplified 
on the front side of the galaxy exposed to the incoming ICM.
In this case, the ICM magnetic field bends along the sides of the galaxy 
and is amplified as it wraps around it. Strong magnetic fields are
also produced as the flow converges behind the galaxy. 
While Case PA produces a flow converging 
from all directions in the $y-z$ plane, Case PP shows 
converging flow only along the $z$-axis, 
resulting in expansion along $y$-axis, i.e., along the direction of
the ICM field.

\begin{figure}
\includegraphics{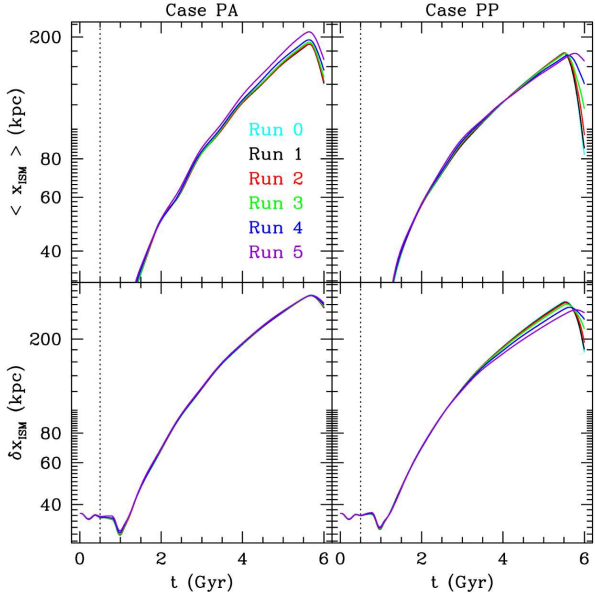}
\caption{Evolution of ${\rm\langle x_{ISM}\rangle}$ 
and ${\rm \delta x_{ISM}}$ in Cases PA ({\it left}) 
and PP ({\it right}). The dotted line corresponds to 0.5 Gyr 
when the ICM inflow begins to enter the simulation box. 
Around 1 Gyr, 
the ISM is strongly compressed by ram pressure, 
resulting in the decrease in ${\rm \delta x_{ISM}}$. 
The tail is slightly less extended in Case PP than in Case PA.
}
\label{fig:morph_x}
\end{figure}

\begin{figure}
\includegraphics{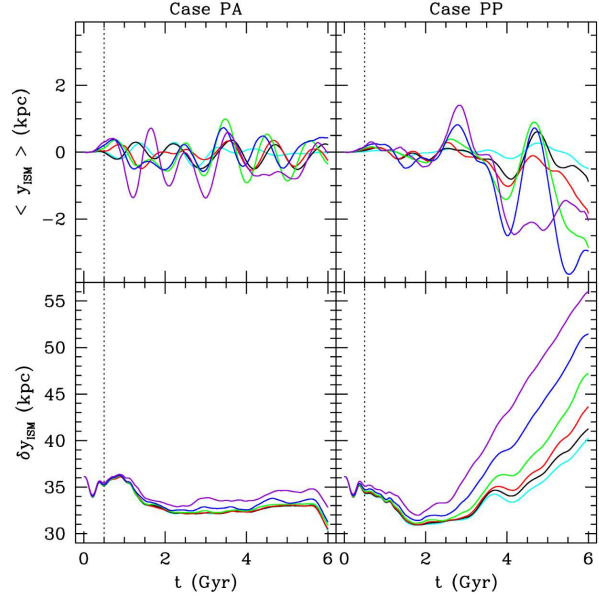}
\caption{Evolution of ${\rm\langle y_{ISM}\rangle}$ 
and ${\rm \delta y_{ISM}}$ in Cases PA ({\it left}) 
and PP ({\it right}). The colour coding of the different lines 
is the same as in Figure \ref{fig:morph_x}. 
In agreement with Figures \ref{fig:3D_para} and 
\ref{fig:3D_perp}, ${\rm \delta y_{ISM}}$ 
increases significantly in Case PP due to 
the expansion of the ISM along the direction 
of the ICM magnetic field. The run with 
the strongest turbulence (i.e., Run 5) shows the largest 
expansion along the $y$-axis.
}
\label{fig:morph_y}
\end{figure}

\begin{figure}
\includegraphics{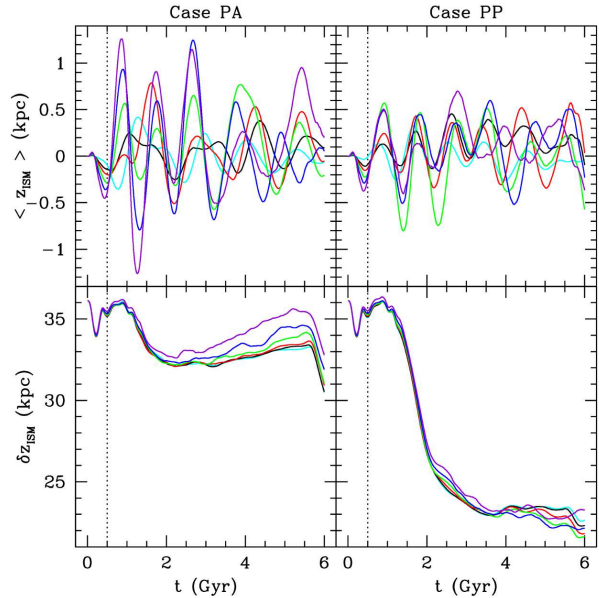}
\caption{Evolution of ${\rm\langle z_{ISM}\rangle}$ 
and ${\rm \delta z_{ISM}}$ in Cases PA ({\it left}) 
and PP ({\it right}). The colour coding of the different lines 
is the same as in Figure \ref{fig:morph_x}. 
In Case PP, as the stripped ISM converges behind the galaxy, 
${\rm \delta z_{ISM}}$ becomes smaller than 
${\rm \delta y_{ISM}}$ at a given time. 
}
\label{fig:morph_z}
\end{figure}

The morphological differences between Case PA and PP are present
even in the early phase of the 
evolution as shown in 
Figures \ref{fig:3D_para} and \ref{fig:3D_perp}. The initial 
stripping is effective in the outer regions of the galaxy. 
In Case PA, a ring-like structure forms 
in the tip of the tail, which is predominantly made of the ISM 
stripped from the outer regions of the galaxy. 
On the other hand, in Case PP the morphology of the stripped ISM is sheet-like. 
The morphological differences grow significantly over time  
as more ISM is stripped from the galaxy. 
Unlike in Case PP, the tail morphology in Case PA more closely resembles that
seen in purely hydrodynamical simulations presented in Paper I.

\subsection{Spatial distribution of the ISM}

We quantify the change in the morphology of the ISM by computing the 
mass-weighted average position and standard deviation of the ISM. For example, 
\begin{equation}
\langle x_{\rm ISM}\rangle = \frac{\sum_{i} C_{i} \rho_{i} V_{i} x_{i}}{\sum_{i} C_{i} \rho_{i} V_{i}},
\label{eq:xavg}
\end{equation}
\begin{equation}
\delta x_{\rm ISM} = \sqrt{ \frac{\sum_{i} C_{i} \rho_{i} V_{i} ( x_{i} -\langle x_{\rm ISM}\rangle )^{2}}{\sum_{i} C_{i} \rho_{i} V_{i}} },
\label{eq:xstd}
\end{equation}
where the index $i$ represents a cell number, 
and $\rho$, $V$, $C$, and $x$, correspond to the density, 
cell volume, colour, and cell $x-$coordinates, respectively.

\begin{figure*}
\includegraphics[scale=0.43]{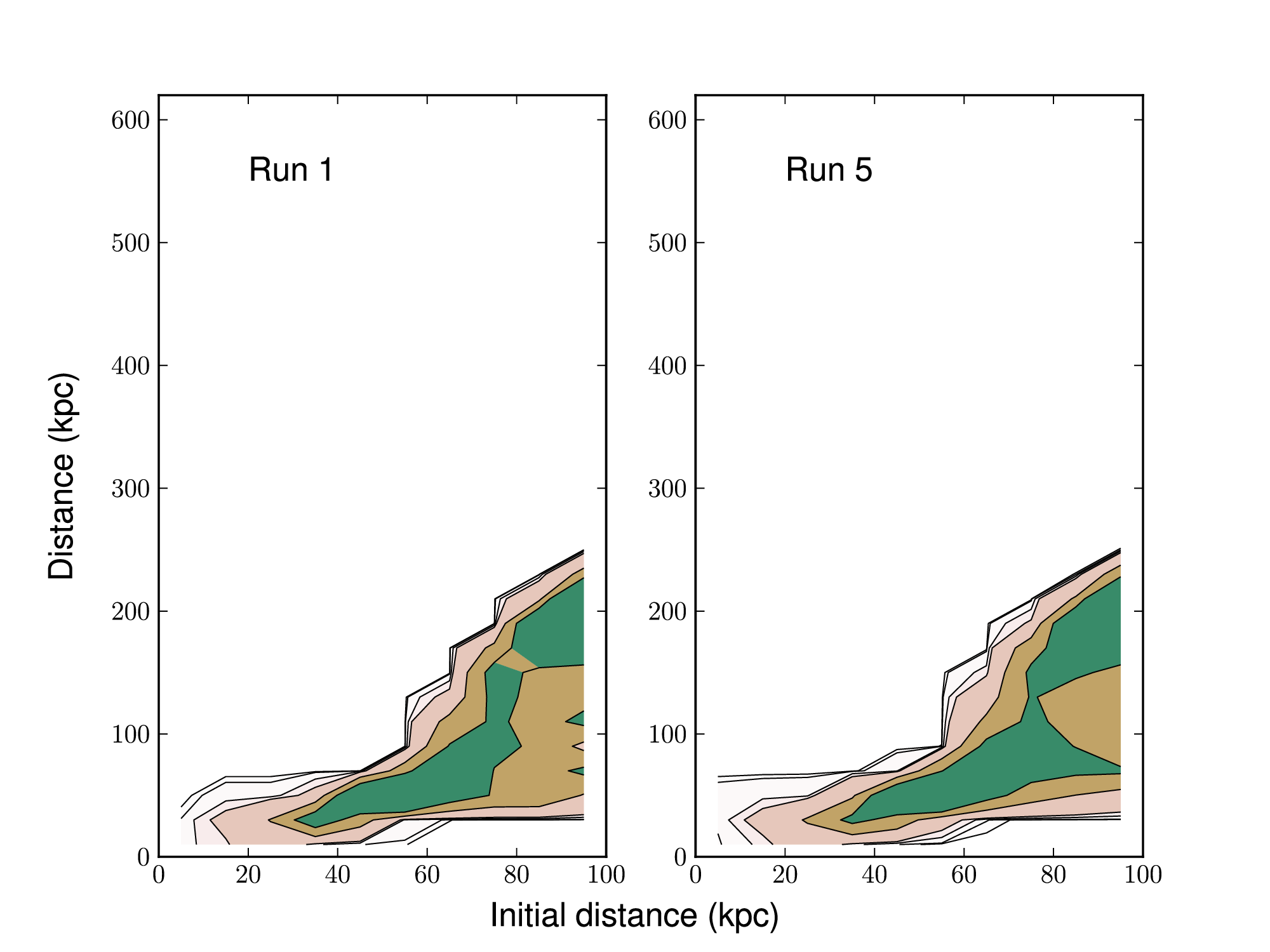}
\includegraphics[scale=0.43]{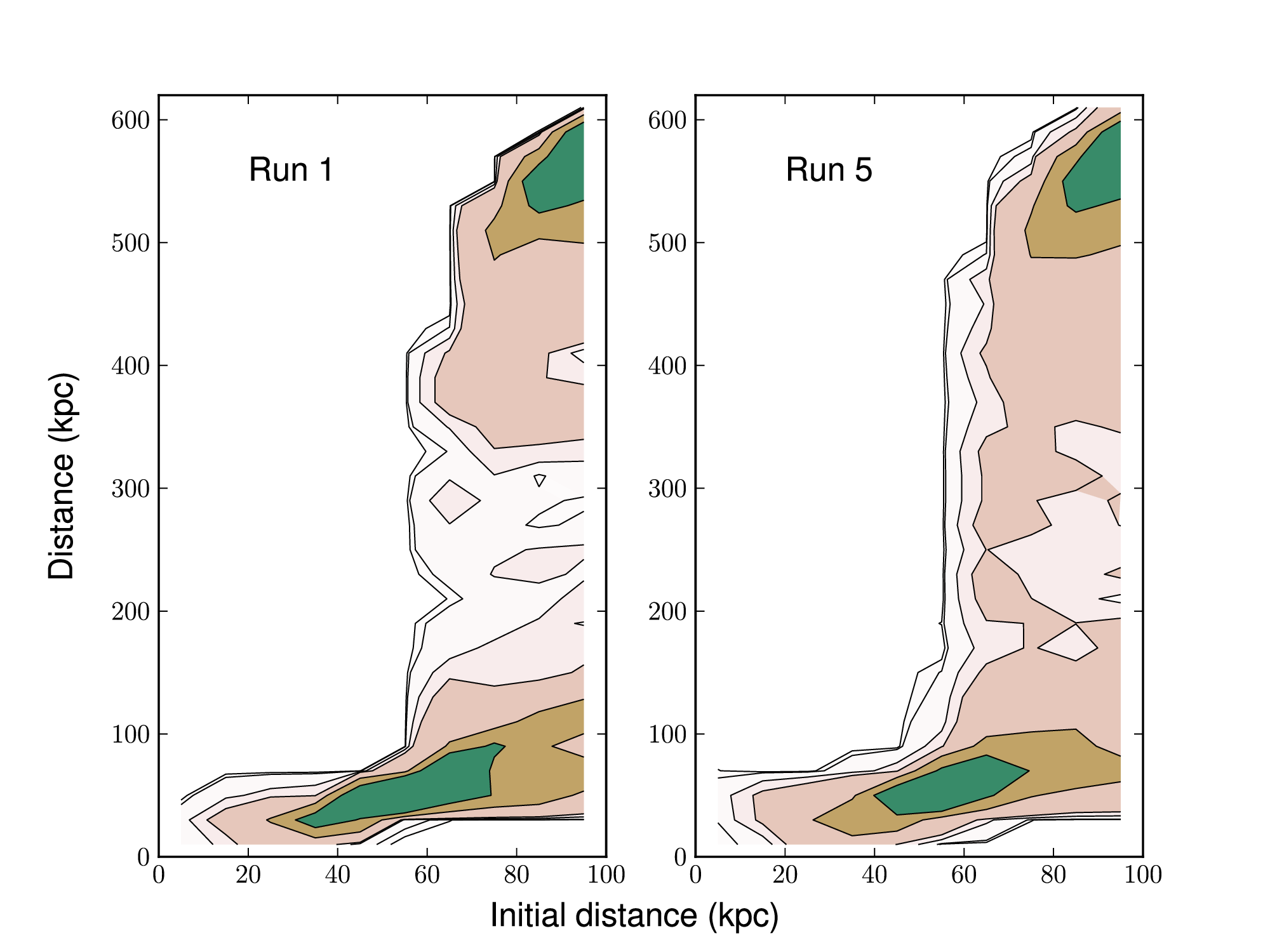}
\caption{Density distribution of the passively moving ISM particles 
in Runs 1 and 5 at 2 ({\it left}) and 4.25 ({\it right}) Gyr in Case PA. 
The number density of the passive particles 
is measured on a uniform grid with bin sizes of 10 
and 20 kpc for horizontal and vertical axes, respectively. 
The colour ranges correspond to bins defined by the number densities 
$5 \times 10^{-5}$, $1 \times 10^{-4}$, 
$5 \times 10^{-4}$, $1 \times 10^{-3}$, 
$5 \times 10^{-3}$, $1 \times 10^{-2}$, 
and $3.5 \times 10^{-2} ~ {\rm kpc^{-2}}$. The distances 
of the particles are measured with respect to the galactic centre.
}
\label{fig:particle_para}
\end{figure*}

\begin{figure*}
\includegraphics[scale=0.43]{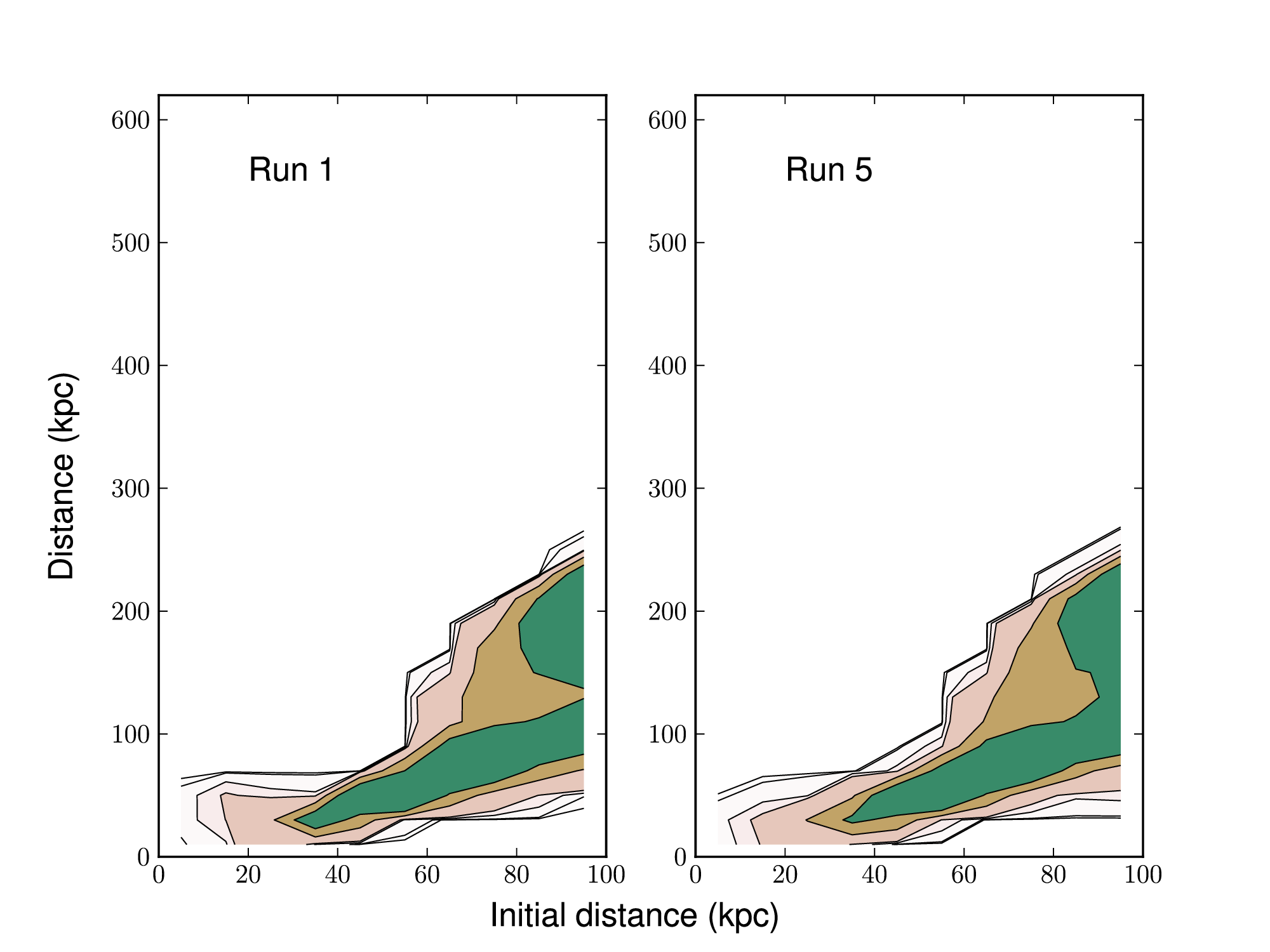}
\includegraphics[scale=0.43]{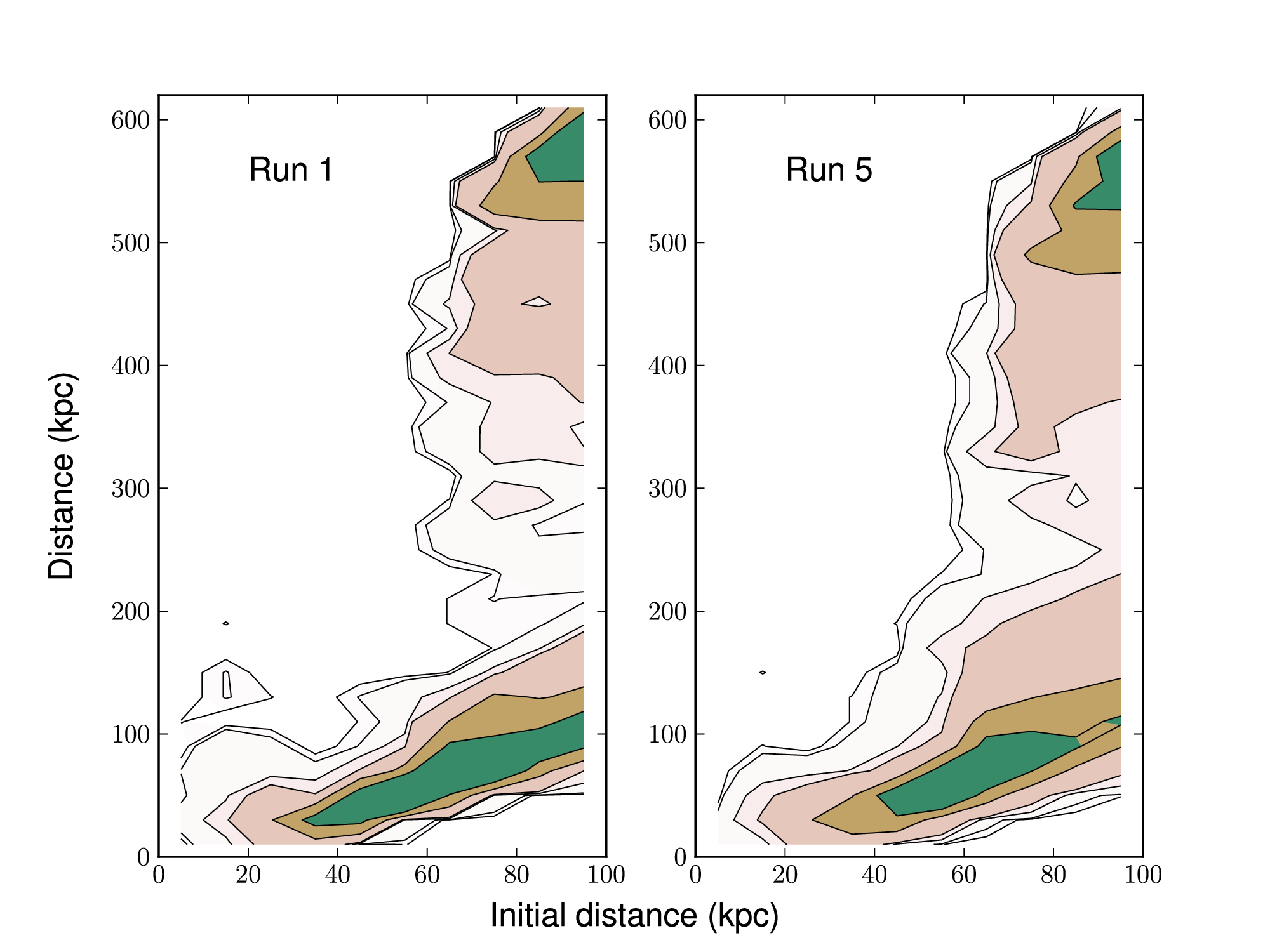}
\caption{Density distributions of the passively moving ISM particles 
in Runs 1 and 5 at 2 ({\it left}) and 4.25 ({\it right}) Gyr 
for Case PP. The format of 
the plots is the same as in Figure \ref{fig:particle_para}.
}
\label{fig:particle_perp}
\end{figure*}

Since the ram pressure acts along the $x$-axis, 
we find the largest displacement of the ISM along that axis 
(see Figure \ref{fig:morph_x}). 
This figure also shows the initial compression 
of the gas around 1 Gyr when the ICM inflow first comes into contact 
with the ISM. The decrease in ${\rm\langle x_{ISM}\rangle}$ 
and ${\rm \delta x_{ISM}}$ 
occurs around 5.5 Gyr because the ISM stripped in the early stage, 
i.e., the ring-like structure seen in Case PA, leaves
the simulation box. 
In both Cases PA and PP, Run 3 leads to longer and wider stripping tail 
compared to the purely hydrodynamic case of Run 2 in Paper I. 
All of these runs have comparable turbulent gas velocity dispersions.
We point out that the dependence of $\langle x_{\rm ISM}\rangle$ and
$\delta x_{\rm ISM}$ on the gas velocity dispersion is much weaker in
the MHD runs compared to pure hydro runs presented in Paper I. 

Figure \ref{fig:morph_y} shows that 
the geometry of the ICM magnetic field strongly alters 
the spatial distribution of the stripped ISM along the $y$-axis. 
Differences appear even in the early phase of stripping. Around 1 Gyr, 
when the initial compression reduces ${\rm \delta x_{ISM}}$ 
in Case PA, ${\rm \delta y_{ISM}}$ increases slightly 
and shows a weak initial expansion along the direction 
perpendicular to the ram pressure direction. This weak expansion is
not seen in Case PP. Instead, in Case PP we observe a 
decrease in ${\rm \delta y_{ISM}}$ up until 2 Gyr. 
This is caused by the stripping of the outer ISM. 
During this stage, the stripped ISM and the magnetic fields, that
bend around the galaxy, are pushed closer 
to the $x$-axis, thus reducing ${\rm \delta y_{ISM}}$.
In Case PA, after 2 Gyr from the beginning of the simulation, 
the ISM in the tail does not change its width 
until around 5.5 Gyr, which marks the moment 
when the tail begins to escape the simulation box. 
On the other hand, in Case PP, the stripped ISM continues to 
spread along the $y$-axis after 2 Gyr. 

The distribution of the gas along the $z$-axis presented 
in Figure \ref{fig:morph_z} also reveals the impact of 
the ICM magnetic field. Because of the symmetry of the field in Case PA, 
the evolution of ${\rm \delta z_{ISM}}$ is 
similar to that of ${\rm \delta y_{ISM}}$. 
In Case PP, as the stripped ISM flows behind the galaxy, 
the gas flow converges along the $z$-axis. This
results in the decrease in 
${\rm \delta z_{ISM}}$ (see Figure \ref{fig:3D_perp}).

The effects of different strength of the ISM turbulence, 
which is quantified in terms of the 1D RMS Mach number (see Table \ref{tab:run}), 
are clearly seen in the evolution 
of ${\rm \delta y_{ISM}}$ in Case PP. However, in Case PA, 
${\rm\langle x_{ISM}\rangle}$, ${\rm \delta y_{ISM}}$, 
and ${\rm \delta z_{ISM}}$ reveal only a weak dependence 
on the turbulence strength. 
As the turbulence strength increases, the tail becomes longer 
and wider in Case PA, but it expands only along $y$-axis in Case PP. 
The trends observed in Case PA are consistent with 
the conclusions from purely hydrodynamic 
simulations (see Figure 2 in Paper I).

\begin{figure*}
\includegraphics{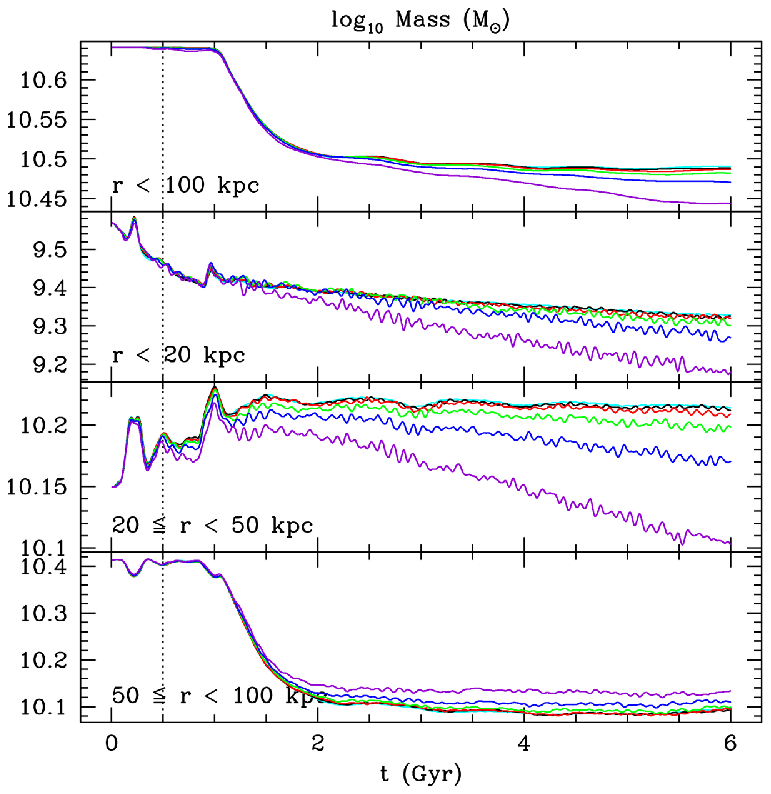}
\includegraphics{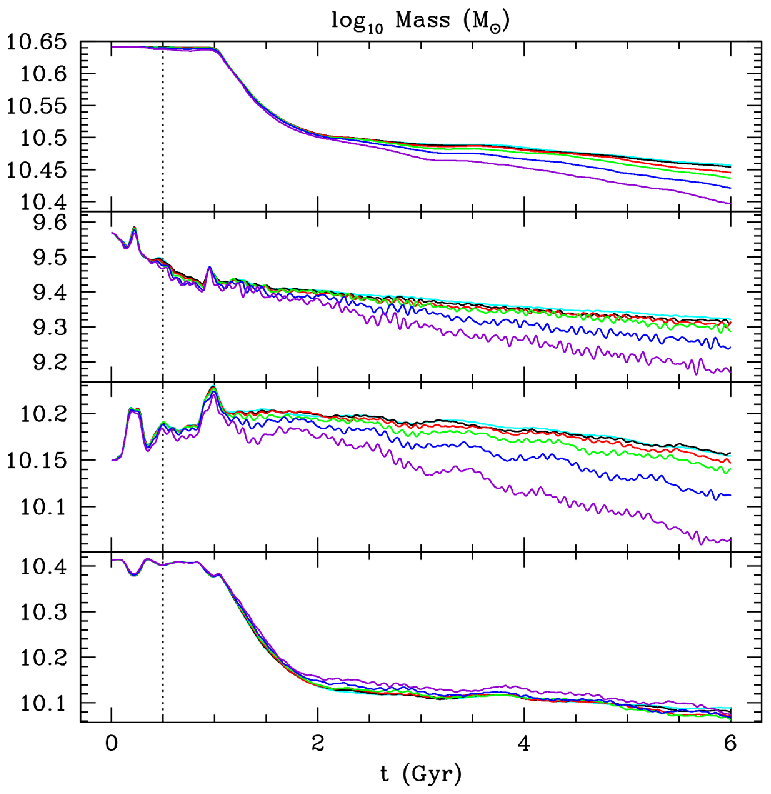}
\caption{Evolution of the mass of intrinsic ISM 
in Case PA ({\it left}) and Case PP ({\it right}). 
From top to bottom, each panel shows the mass 
in four different radial zones:  $r < 100$ kpc, 
$r < 20$ kpc, 20 kpc $\leq r <$ 50 kpc, and 50 kpc $\leq r <$ 100 kpc 
($R_{t}=100$ kpc). 
The colour coding of the different lines 
is the same as in Figure \ref{fig:morph_x}.
The dotted vertical line 
corresponds to 0.5 Gyr when the ICM flow begins to enter the simulation box. 
Note that the vertical axis range is different in each panel.
}
\label{fig:mass_ISM}
\end{figure*}

\subsection{Origin of the ISM in the tails}

We trace the origin of the ISM in the stripping tail 
by examining the distributions of 
the passively moving ISM particles. These particles 
are initially distributed uniformly
inside $R_{t}$. As the stirring process progresses, 
the ISM particles located initially in the galactic 
centre travel to outer regions 
of the galaxy due to random turbulent gas motions. 
These particles can be subsequently removed by ram pressure and
transported to the tail.

Figure \ref{fig:particle_para} shows 
the spatial distribution of the ISM particles 
in Case PA. While in the initial stages (at 2 Gyr) 
the stripping of particles originally located at $r {\rm ~ > ~
70}$ kpc is just as effective in Run 1 and 5, 
at 4.5 Gyr the distribution of particles on the 
distance versus initial distance plane becomes 
thicker, i.e., more ISM particles originate from initial distances $r
{\rm ~ > ~50}$ kpc, in Run 5 than Run 1. 
This difference is visible in Figure 
\ref{fig:particle_para} where the effective area corresponding to the 
particle densities ranging from $1\times10^{-3}$ to 
$5\times10^{-3} {\rm kpc^{-2}}$ is larger in Run 5 than in Run 1 for 
final distances greater than 100 kpc.

As shown in Figure \ref{fig:particle_perp},
the tail in Case PP is weaker and 
consists of the ISM from a slightly narrower 
spatial range of initial distances than in Case PA. 
This difference is caused by less efficient stripping of the ISM 
originally located in the range $r ~ > ~ 60$ kpc 
in Case PP compared to Case PA.
Even though Run 5 in Case PP shows turbulence-enhanced ISM loss 
over this spatial range when compared to Run 1 in Case PP, 
it still produces weaker tail than Run 5 in Case PA. 
This manifests itself as the 
difference in the spatial distribution of particle densities 
ranging from $1\times10^{-3}$ to 
$5\times10^{-3} {\rm kpc^{-2}}$ 
in the distant parts of the tail.

We note that the differences between Case PA and PP
are not due to turbulence strengths. Table \ref{tab:run} shows that
the runs have 
very similar gas velocity dispersions in Cases PA and PP. Therefore, we
attribute the differences in the morphology of the distributions shown
in Figures \ref{fig:particle_para} and \ref{fig:particle_perp} 
to the magnetic field configurations rather than turbulence strengths.

\subsection{Evolution of the ISM mass retained in the galaxy}

The decrease of the ISM mass as a function of time for different
radial zones is shown in Figure \ref{fig:mass_ISM}. 
The largest amount of stripping is found in the runs with the
strongest turbulence, i.e., Run 5. 
The galaxy in Case PA retains slightly larger amount of the ISM in the outer 
regions ($50\la r\la 100$ kpc) than in Case PP. 
Efficient mixing leads to the loss 
of the ISM that was originally located in the centre ($r < 50$ kpc). 
Since this effect is weaker in Run 0, 
the galaxy in Run 0 retains about 11\% more ISM 
than Run 5 at 6 Gyr in Case PA. 

Considering only the effect of the magnetic fields on the ISM mass loss, 
we find that the magnetic field suppresses mass loss rates 
compared to the pure hydrodynamic cases. 
For example, at 6 Gyr the ISM mass 
remaining inside the galaxy in Run 3 is larger  
than that in a purely hydrodynamic simulation (i.e., Run 2 presented in
Paper I). This choice of comparison runs is meaningful because 
we are comparing two different simulations characterised by a very
similar level of the ISM turbulence as described by the magnitude of 
1D RMS velocity dispersions (see Table 1 in Paper I and here). 
Specifically, in this example, in the MHD run of Case PA 
the galaxy retains 16\% 
more ISM despite the fact that the 1D RMS velocity dispersion in that run 
is similar to that of the hydro Run 2 in Paper I.

The geometry of the ICM magnetic field also 
affects the efficiency of the ISM stripping. 
These differences are illustrated in Figure \ref{fig:mass_ISM}. 
For example, the galaxy in Run 1 in Case PA retains about 8\% more ISM 
at 6 Gyr than Run 1 in Case PP despite the similar strength of turbulence 
in the beginning of the simulations although this 
difference in the ISM masses is only about 1\% at 3 Gyr.
The comparison of Run 5 results at 6 Gyr reveals that 
the galaxy in Case PA retains about 
10\% more ISM than in Case PP.

Considering the effects of both the magnetic field and 
the ISM turbulence together, we find that, irrespectively of the
magnetic field orientation, the very presence of the magnetic field 
leads to a smaller spread in the ISM loss rates for a given spread in the 1D
RMS velocity dispersion. For example, at 6 Gyr, in Run 0 of Case PP 
the galaxy retains 
about 15\% more ISM than in Run 5. This difference is smaller than the
one between Run 0 and Run 3 in the purely hydrodynamic case presented 
in Paper I, 
despite the fact that the gas velocity dispersions in this MHD case is
even larger than that in the hydro case.

\begin{figure*}
\includegraphics{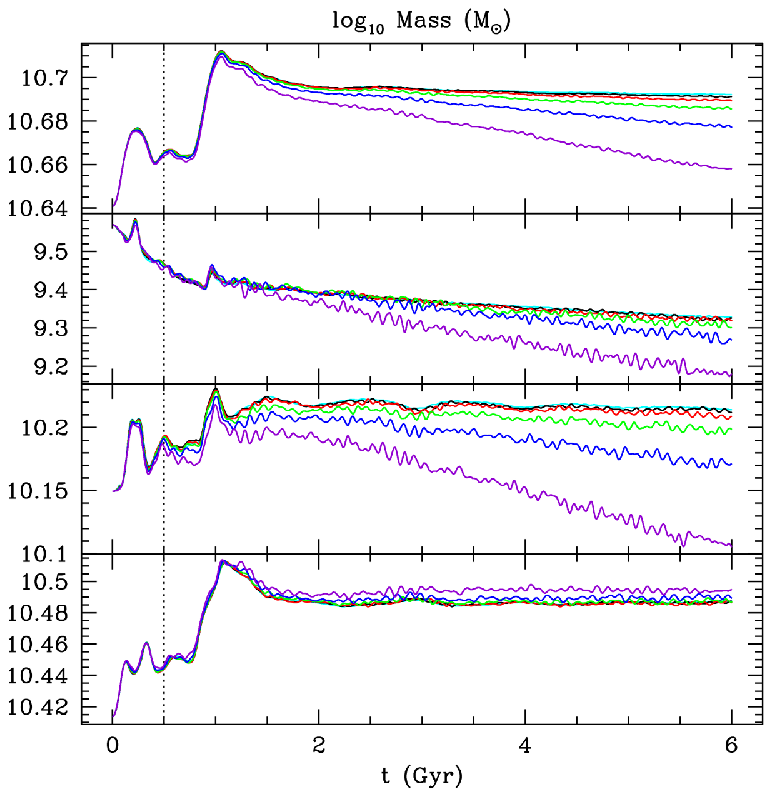}
\includegraphics{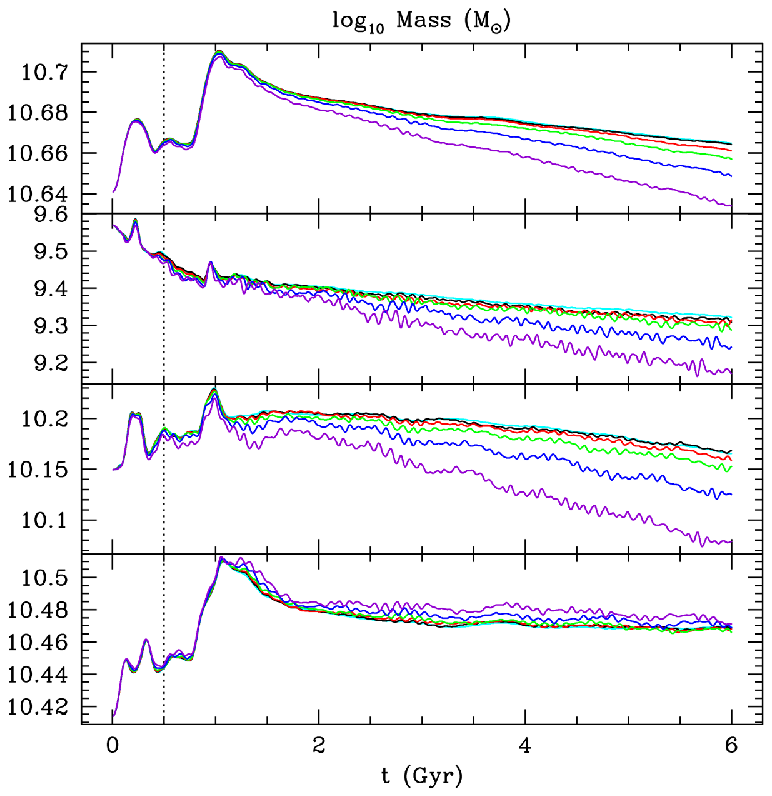}
\caption{Mass evolution of the total gas mass 
inside $R_{t}$ for Cases PA ({\it left}) and PP ({\it right}). 
From top to bottom, each panel shows the mass 
in four different radial zones: $r < 100$ kpc, 
$r < 20$ kpc, 20 kpc $\leq r <$ 50 kpc, and 50 kpc $\leq r <$ 100 kpc 
($R_{t}=100$ kpc). The dotted vertical line corresponds to 0.5 Gyr 
when the ICM starts to enter the simulation box. 
The colour coding of the different lines is the same 
as in Figure \ref{fig:morph_x}. Note that the ranges of the vertical axes 
are different in the left and right columns.
}
\label{fig:mass_gas}
\end{figure*}

\begin{figure*}
\includegraphics[scale=0.43]{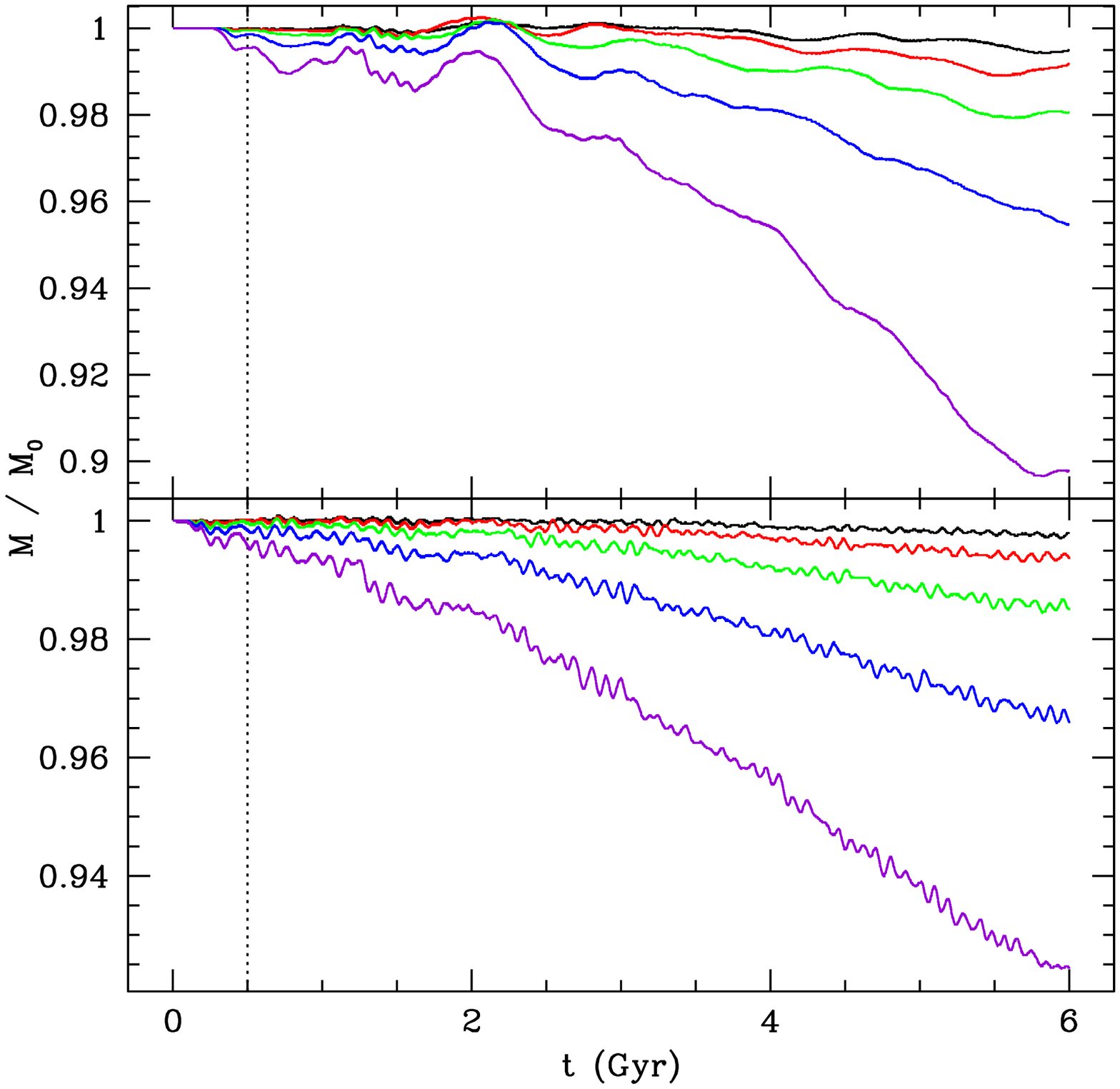}
\includegraphics[scale=0.43]{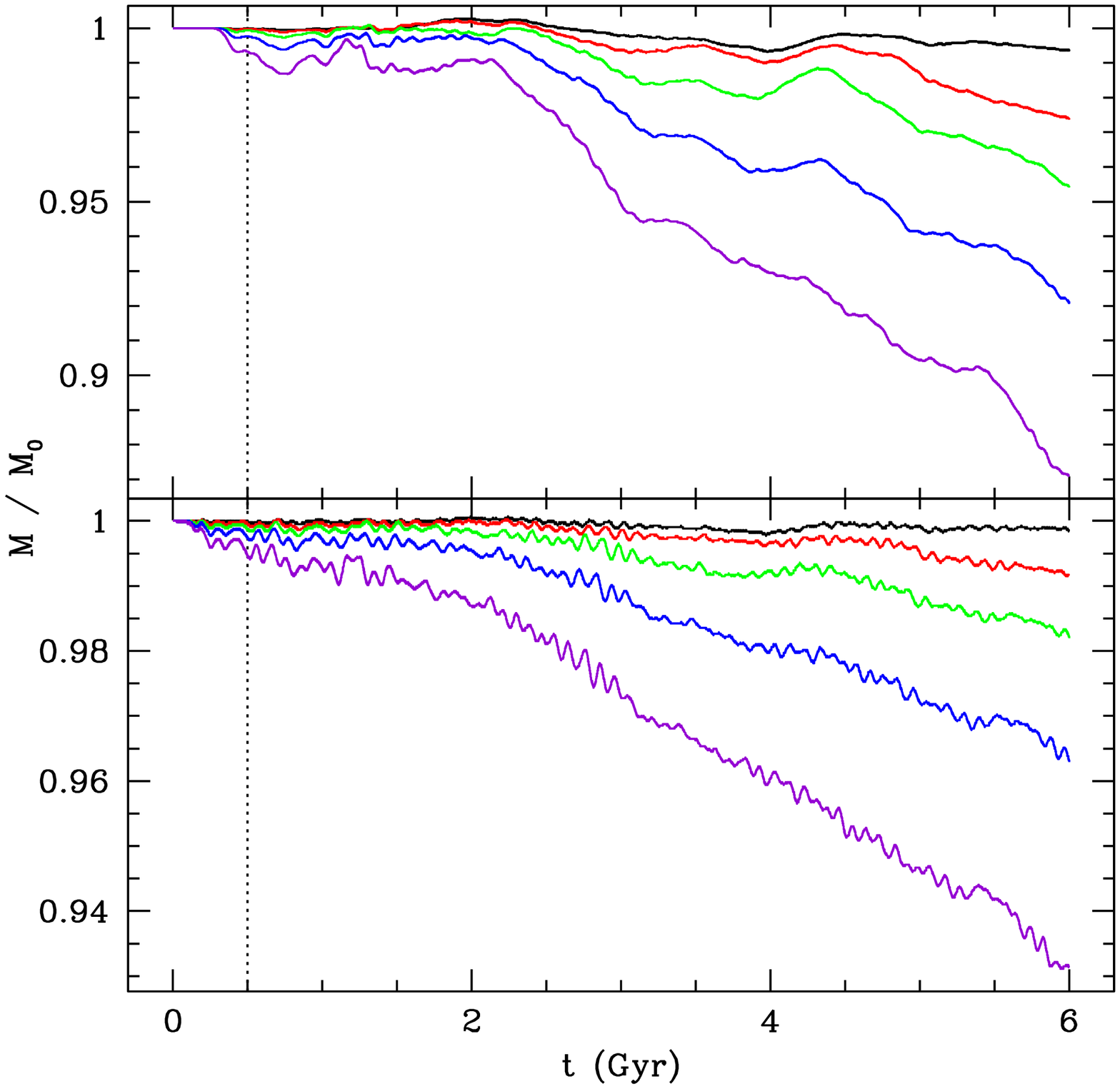}
\caption{Evolution of the intrinsic ISM mass ({\it top}) and the total
gas mass ({\it bottom}) inside $R_{t}$ 
with respect to mass evolution of Run 0 in 
Cases PA ({\it left}) and PP ({\it right}). 
The colour scheme is the same as in Figure \ref{fig:morph_x}. 
The dotted vertical line corresponds to 0.5 Gyr 
when the ICM starts to flow into the simulation box.}
\label{fig:ICM_over_ISM}
\end{figure*}

In summary, we find that increasing the level 
of turbulence in the ISM enhances the ISM loss. 
However, the strength of 
the ISM turbulence, as 
quantified by the 1D RMS velocity dispersion here, 
affects the ISM mass loss to lesser extent in the MHD simulations than
in pure hydrodynamic ones. 
The geometry of the magnetic field in the ambient 
ICM alters the ISM loss rates.

\begin{figure*}
\includegraphics[scale=0.43]{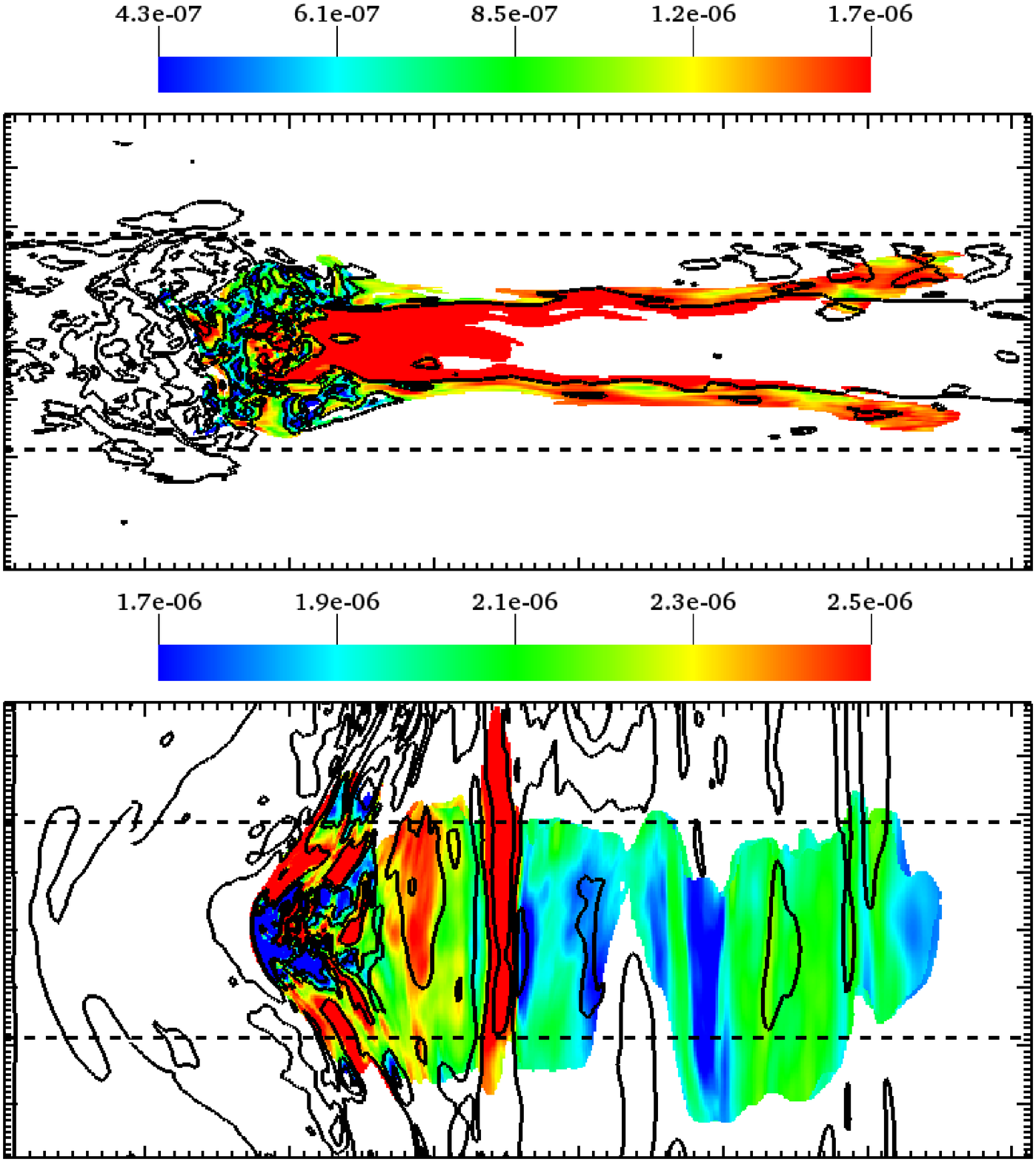}
\caption{Cross-section through the magnetic-field magnitude 
distribution in Run 5 on $x-y$ plane 
centred on the galactic centre. The snapshot is taken at 5 Gyr.
Results for Case PA are shown in the top panel and for Case PP in the
bottom panel. 
Contour lines correspond to linearly-spaced values of the magnetic
field magnitude ranging from  
$4.35 \times 10^{-7}$ to $1.65 \times 10^{-6}$ (G), 
and from $1.7 \times 10^{-6}$ to $2.5 \times 10^{-6}$ (G), in  
Cases PA and PP, respectively. The coloured regions 
represent areas with ISM density above 1\% of total 
(i.e., ICM plus ISM) gas density. The 
unit of the colour bar is Gauss. The dashed lines 
correspond to 120 kpc from the galactic centre along the $x$-axis.
}
\label{fig:B_field}
\end{figure*}

\subsection{Evolution of the total gas mass inside the galaxy}

The inflowing ICM replaces the ISM inside $R_{t}$ over time, 
and it blends with the ISM remaining in the galaxy. 
Figure \ref{fig:mass_gas} shows that
the total gas mass inside $R_{t}$ increases 
in both Case PA and PP early on ($t < 1$ Gyr), while the total amount 
of the ISM changes little (see Figure \ref{fig:mass_ISM}). 
This increase is mainly found in the outer region 
of the galaxy ($r<$50 kpc). 
At later times, the total mass of the gas continuously decreases 
because the ICM is captured only temporarily in 
the outer region of the galaxy. Both the ISM and ICM are removed from
the galaxy due to the combined action of turbulence, mixing, and ram 
pressure stripping. In analogy to the ISM mass stripping inside $R_{t}$, 
we find that more total mass is removed 
from the galaxy in Case PP than in Case PA.

\begin{figure*}
\includegraphics[scale=0.43]{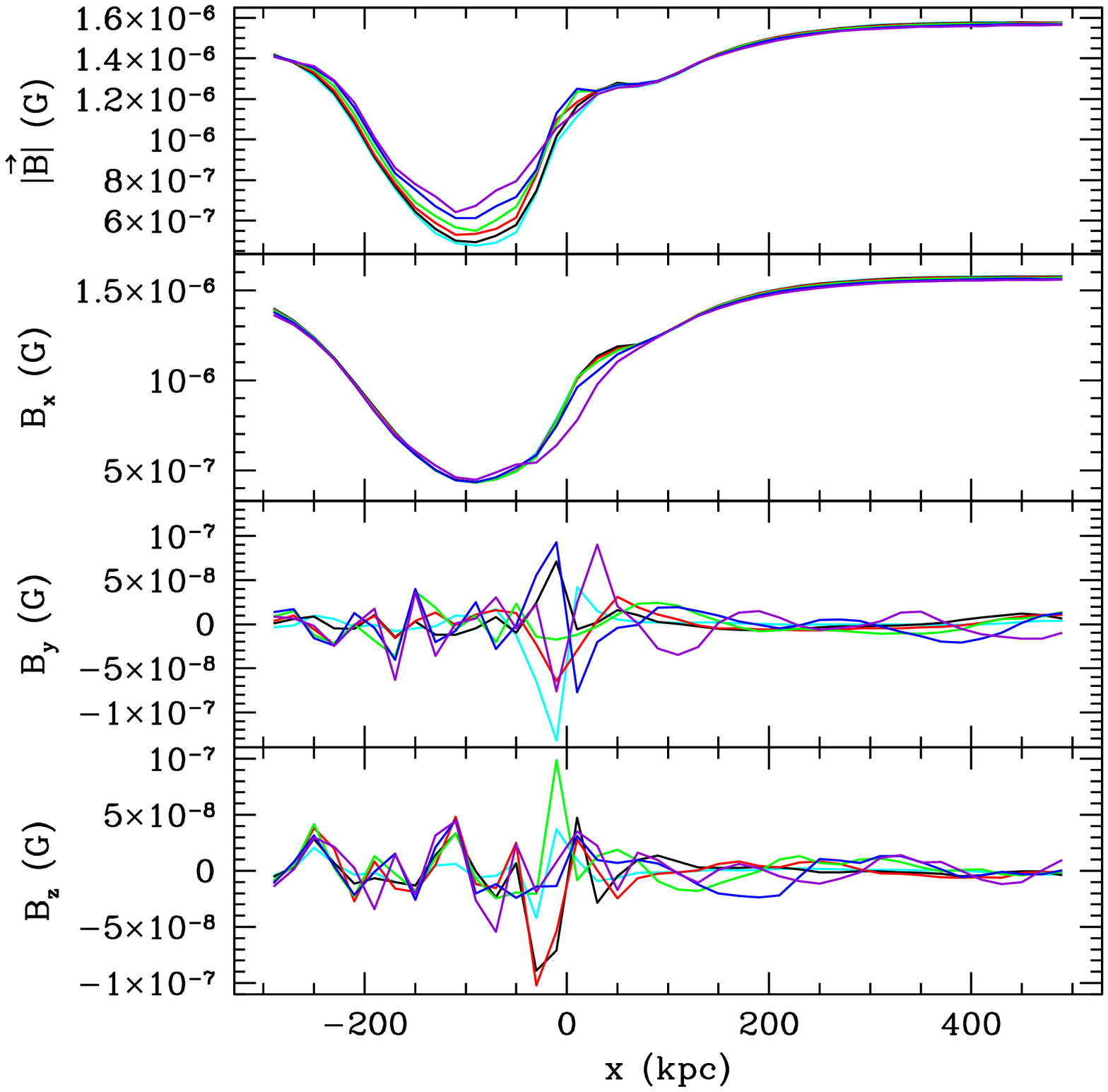}
\includegraphics[scale=0.43]{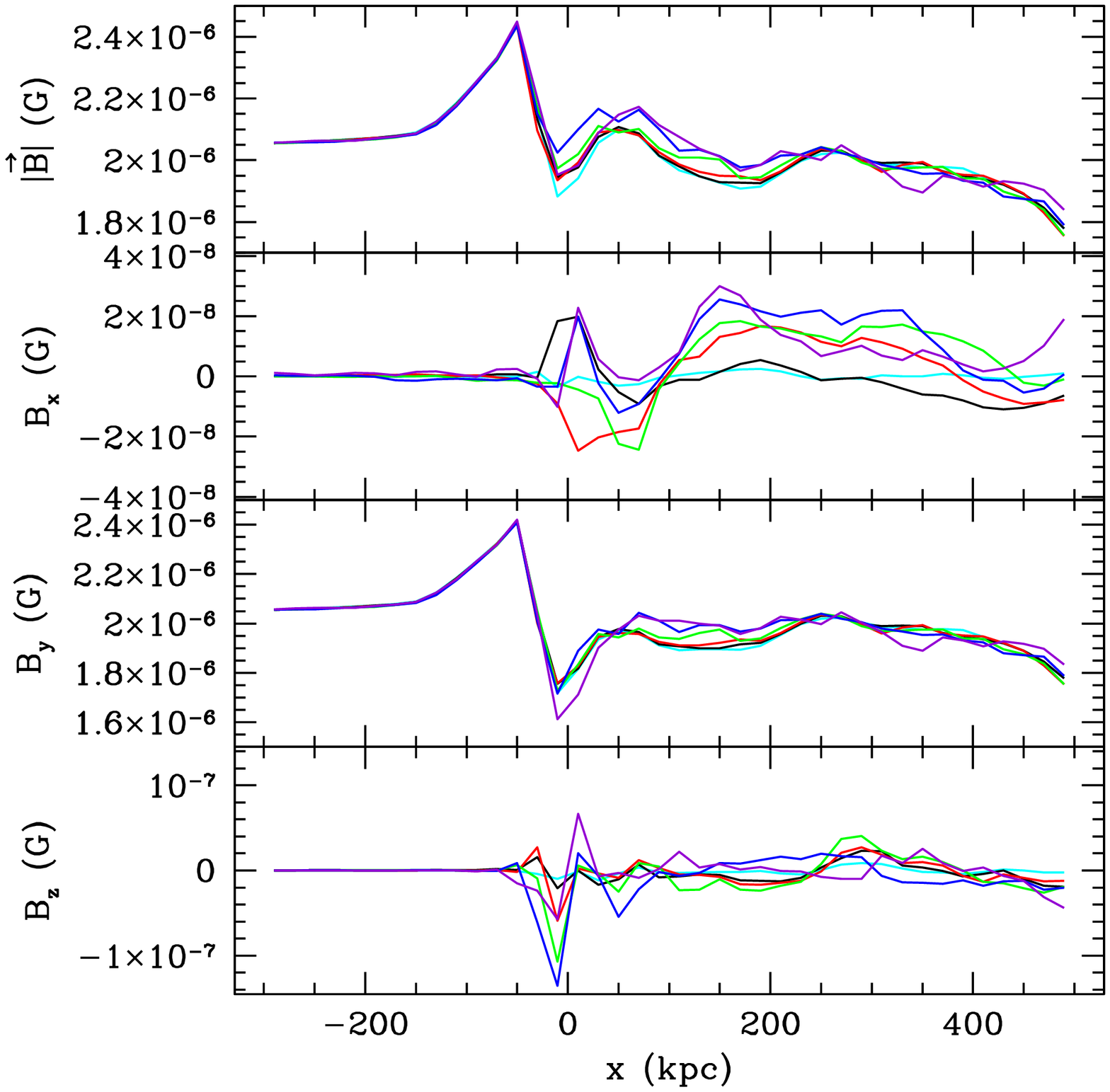}
\caption{Mass-weighted magnetic field magnitude, 
$x$, $y$, and $z$-components of the magnetic field (from top to bottom) 
along the $x$-axis inside a cylinder 
of radius 120 kpc at 5 Gyr for Cases PA ({\it left}) and PP ({\it right}). 
The bin size along the $x$-axis is 20 kpc. 
The colour coding of the different lines is the same 
as in Figure \ref{fig:morph_x}.
}
\label{fig:B_field_profile}
\end{figure*}

\begin{figure*}
\includegraphics[scale=0.43]{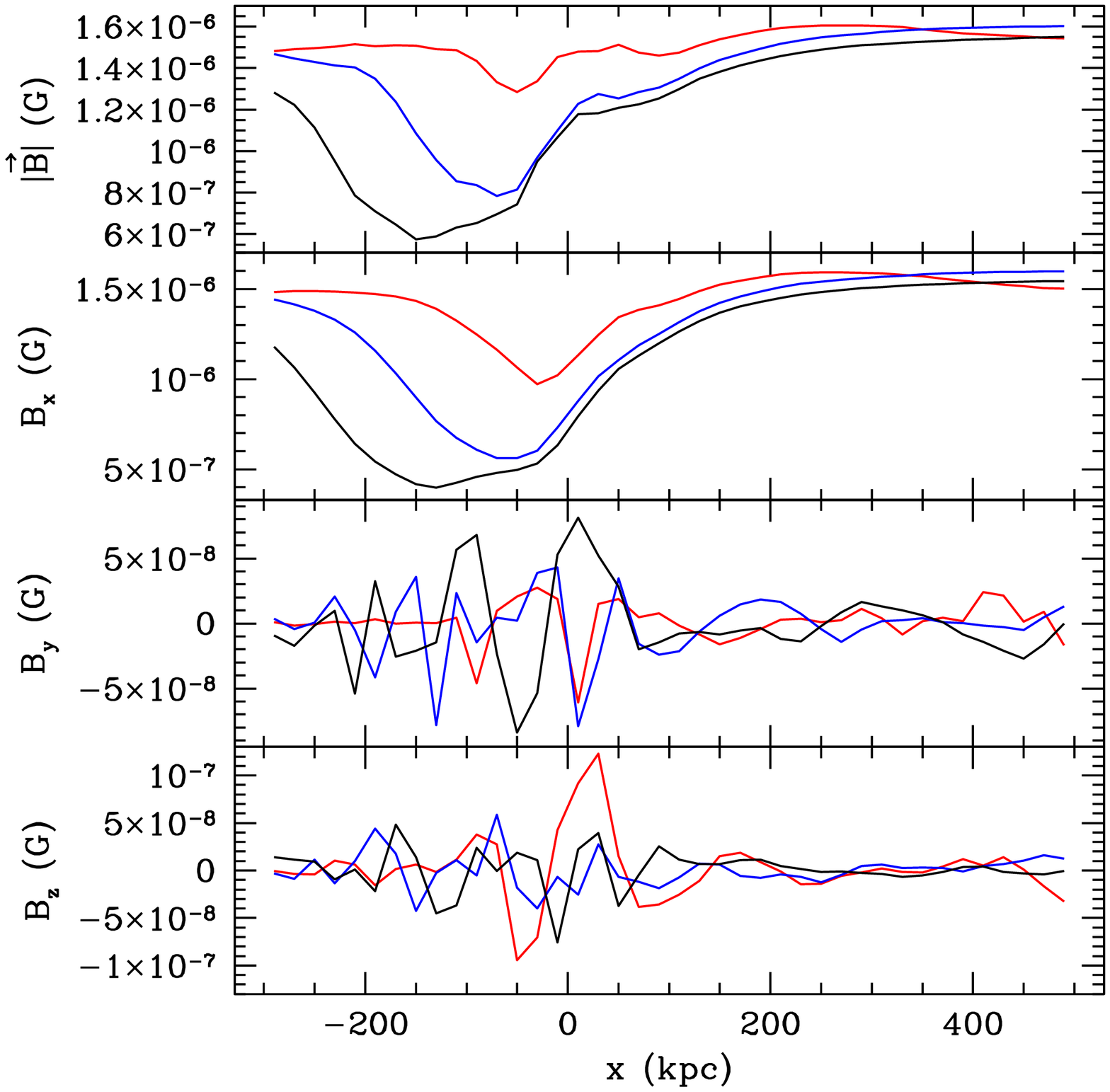}
\includegraphics[scale=0.43]{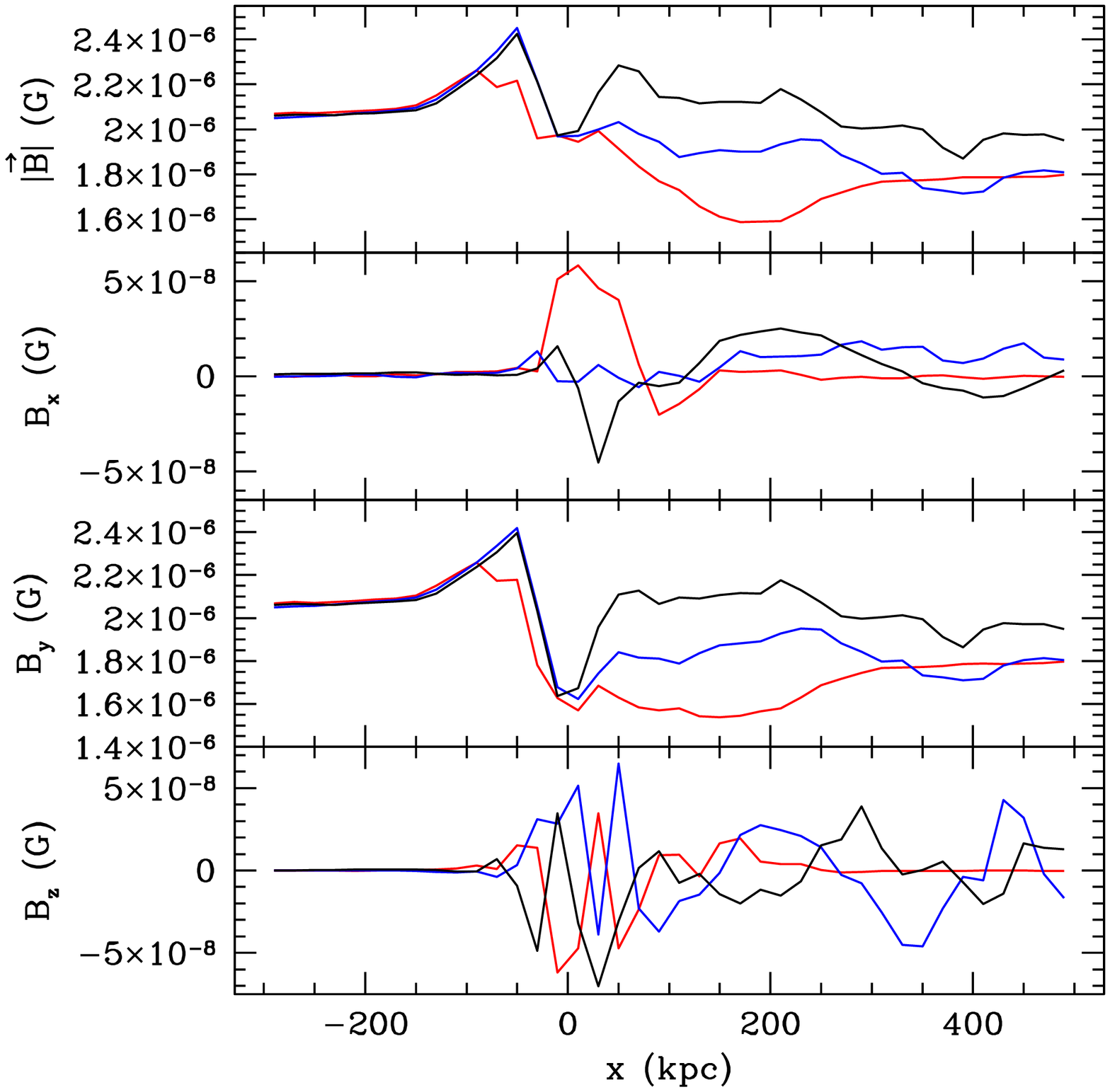}
\caption{Mass-weighted magnetic field magnitude,  
$x$, $y$, and $z$-components of the magnetic field (from top to bottom)
along the $x$-axis inside a cylinder 
with a radius 120 kpc in Run 5 for Cases PA ({\it left}) and PP ({\it right}) 
at 2 ({\it red}), 4 ({\it blue}), and 6 ({\it black}) Gyr. The bin
size along the $x$-axis is 20 kpc.
}
\label{fig:B_field_time}
\end{figure*}

As Figure \ref{fig:ICM_over_ISM} shows, 
we find that the relative change in the ISM mass is larger than the 
corresponding change in the total gas mass. This figure also shows 
that the variation in the retained ISM mass in Case PP is larger than
in Case PA despite the smaller difference of 1D RMS velocity dispersions 
between Runs 0 and 5 in Case PP than in Case PA (see Table \ref{tab:run}).
However, the total amount of gas inside $R_{t}$ for different
turbulence levels is very similar in Cases PA and PP.

Comparing hydrodynamics runs to MHD runs, 
both MHD Cases PA and PP exhibit less
variation in the ISM mass and total gas mass
with the ISM turbulence strength. For example, at 6 Gyr,
the difference between Run 0 and 3 is about 2\% in Case PA and 4\%
variations in Case PP (see Figure \ref{fig:ICM_over_ISM}). 
Despite the fact that the difference in the ISM turbulence strength
between Run 0 and 2 in the pure hydro case is very similar to that
between Run 0 and 3 (either for Case PA or PP),  
the difference between the ISM mass remaining in the galaxy between
Run 0 and 2 is about 6\%, i.e., larger than in the MHD case described
above. As far as the comparisons between the total gas mass are
concerned, the comparison between the same pairs of simulations as
above shows that the differences in the MHD cases are less than 0.5\%
for both Cases PA and PP, while the differences in the pure hydro
cases are about 2\% (see Figure 7 in Paper I). These comparisons prove that 
the strength of the ISM turbulence affects the ISM mass loss 
to lesser extent in the MHD simulations than 
in pure hydrodynamic ones.

In summary, strong ISM turbulence enhances mixing and mass loss. The
effective total gas mass loss occurs despite the fact that
the ICM replaces the ISM in the galaxy. This effective mass loss is
possible because the ICM is incorporated into the galaxy only temporarily.
However, the effect of turbulence has less impact on the stripping
rates in the MHD simulations than in the hydrodynamic ones. 
When the ambient ICM magnetic field 
is perpendicular to the direction of the ram pressure (i.e., Case PP), 
the total gas mass decreases more rapidly than when the fields are
parallel to the direction of ram pressure (i.e., Case PA).

\subsection{Distribution of the magnetic fields}

The differences in the evolution of the mass stripping in Cases PA and PP
are closely related to the geometry of the magnetic field in these two
cases. There is a marked difference between the magnetic field distributions
in these two cases. Investigating the distribution of the magnetic field in
the galaxy and the stripping tail 
for simplified geometries of the external ICM magnetic fields is the
first step that we need to take in order to correctly interpret the
results of more sophisticated simulations with more realistic field
topologies and make detailed predictions for observable quantities.

\begin{figure*}
\includegraphics{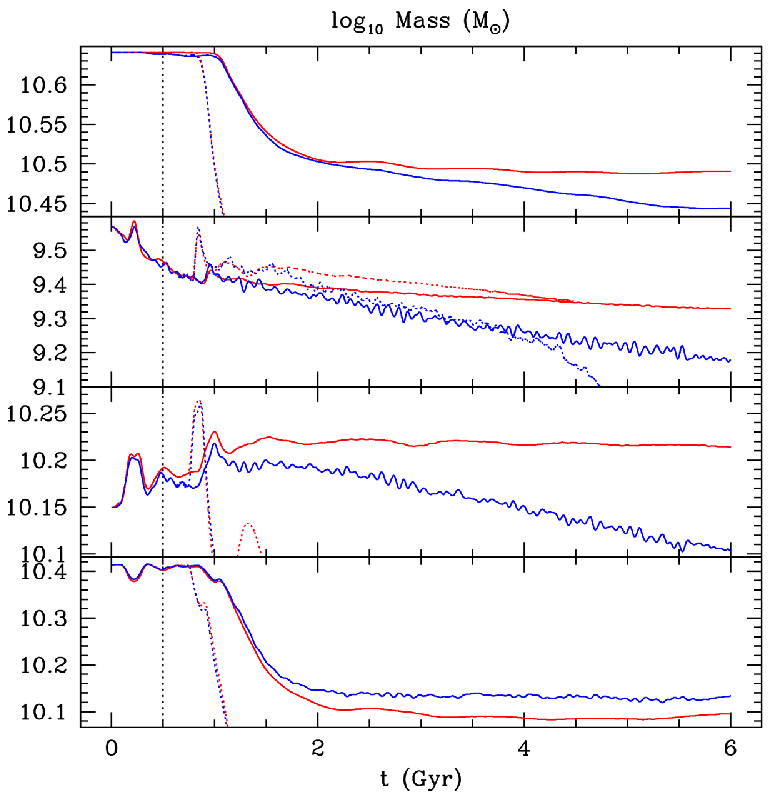}
\includegraphics{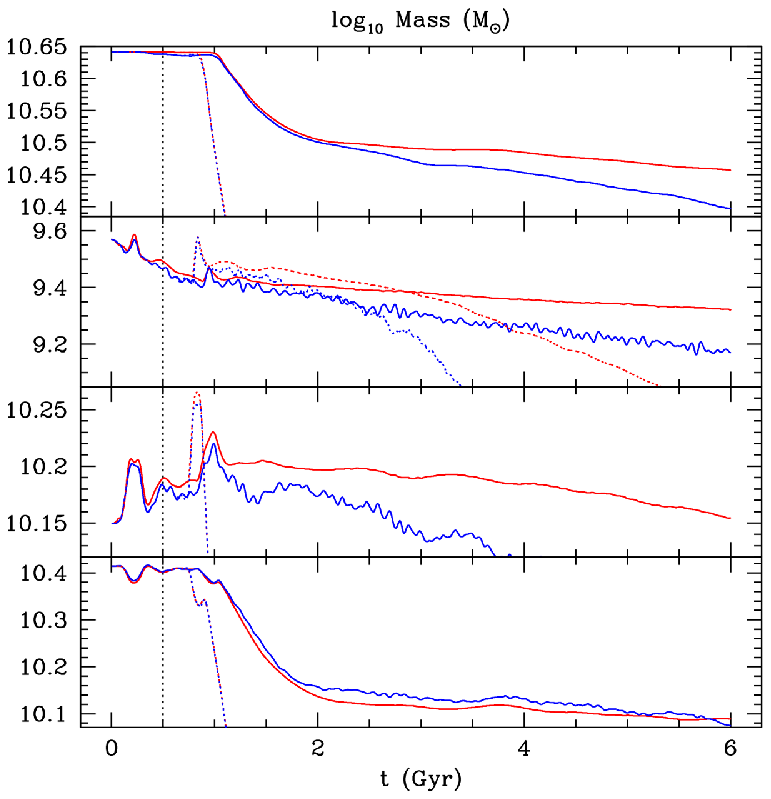}
\caption{Mass evolution of the intrinsic ISM 
for Cases PA ({\it left}) and PP ({\it right}) 
in Runs 0h ({\it red}) and 5h ({\it blue}). 
From top to bottom, each panel shows the mass 
in four different radial zones: $r < 100$ kpc, 
$r < 20$ kpc, 20 kpc $\leq r <$ 50 kpc, and 50 kpc $\leq r <$ 100 kpc 
($R_{t}=100$ kpc).
The solid lines correspond to Runs 0 and 5 and the dotted lines
correspond to Runs 0h and 5h. 
The black dotted line corresponds to 0.5 Gyr 
when the ICM flow begins to enter the simulation box.
}
\label{fig:ISM_high_low_comp}
\end{figure*}

\begin{figure*}
\includegraphics[scale=0.43]{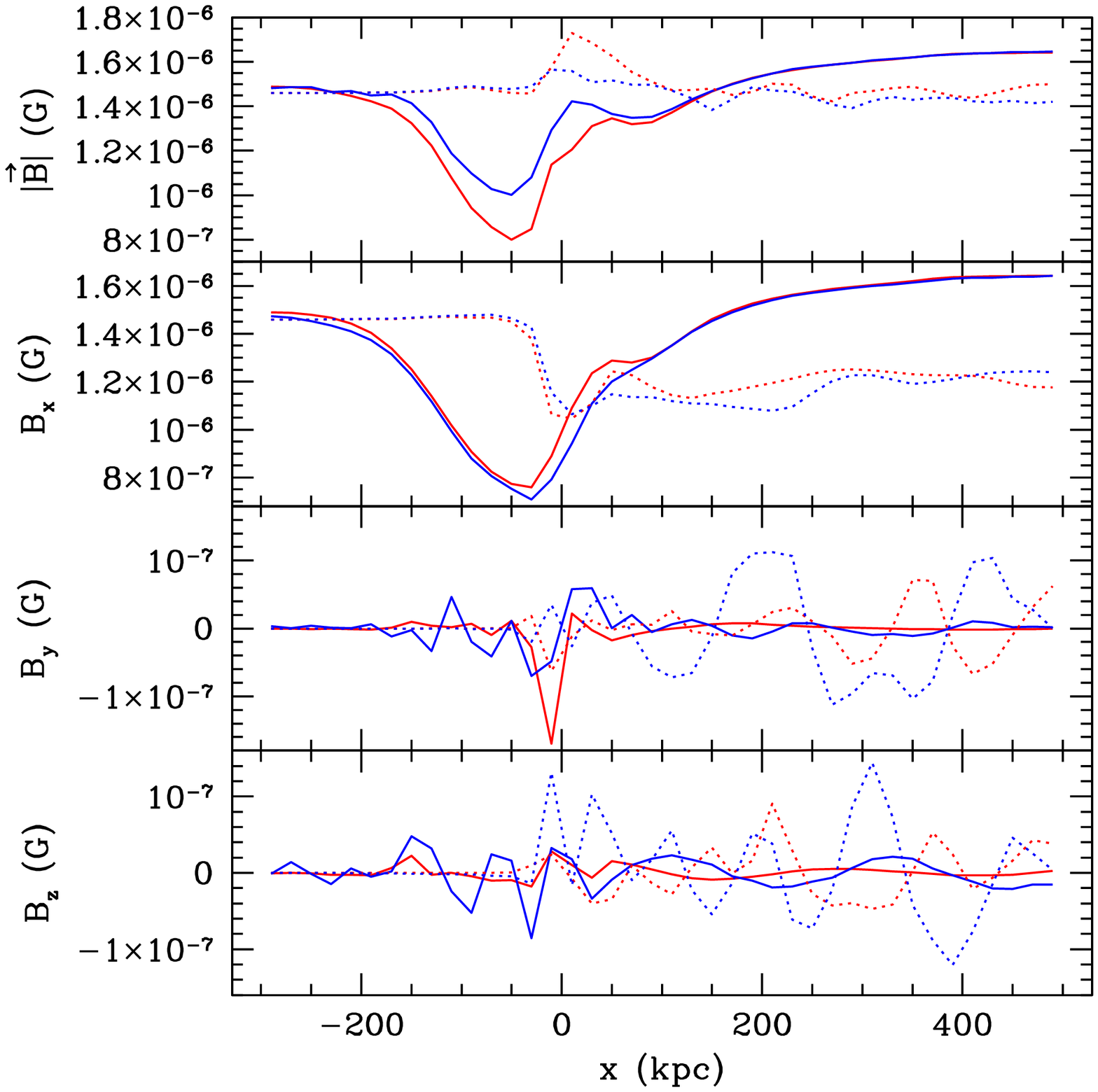}
\includegraphics[scale=0.43]{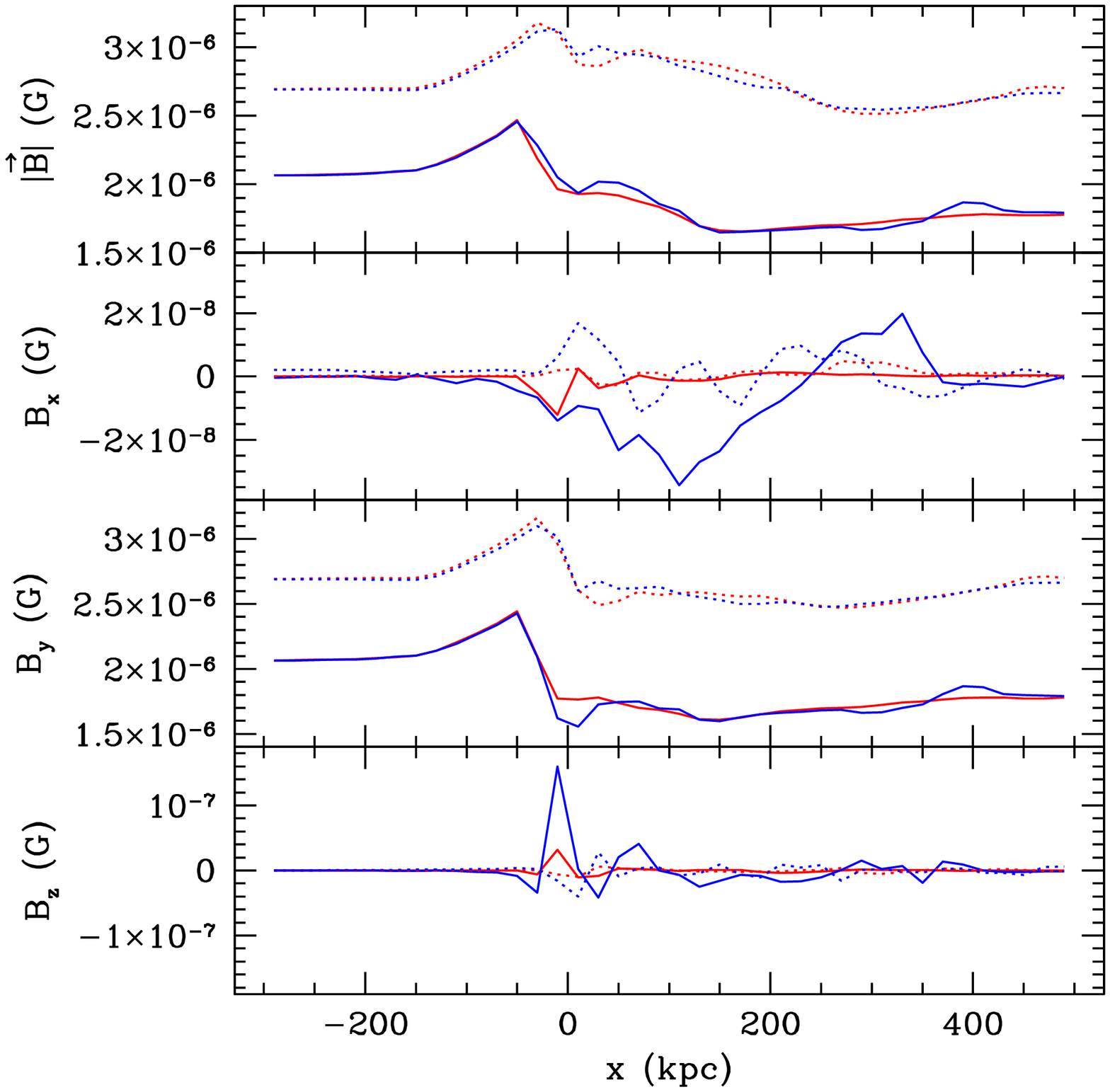}
\caption{Mass-weighted $x$, $y$, and $z$ components of the magnetic field 
(from top to bottom) along the $x$-axis inside a cylinder of 
radius 120 kpc at 3 Gyr for Cases PA ({\it left}) and PP ({\it right}). 
The bin size along the $x$-axis is 20 kpc. 
Colour coding and line styles are the same 
as in Figure \ref{fig:ISM_high_low_comp}. 
}
\label{fig:B_high_low_comp}
\end{figure*}

\begin{figure*}
\includegraphics[scale=0.217]{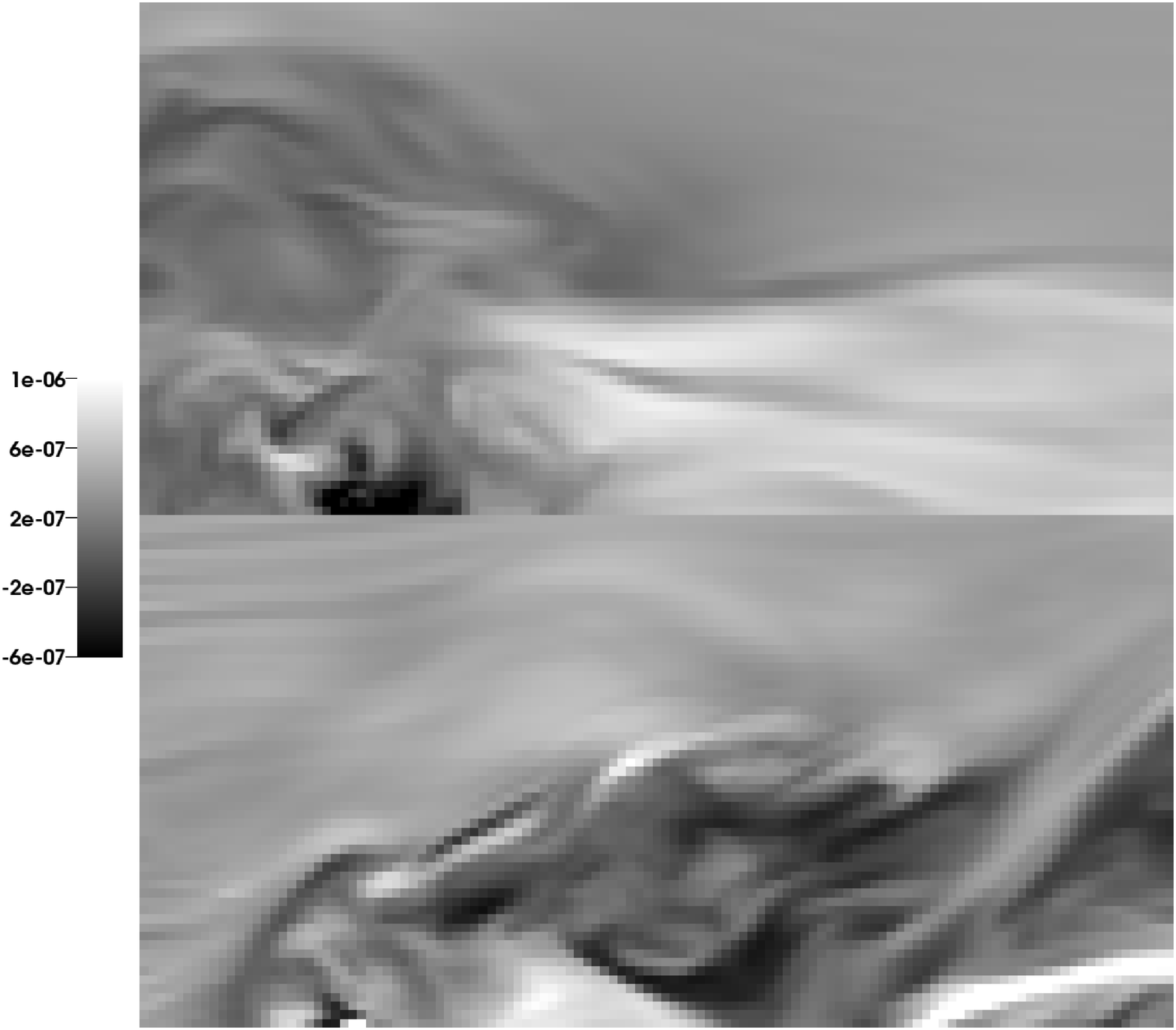}
\includegraphics[scale=0.217]{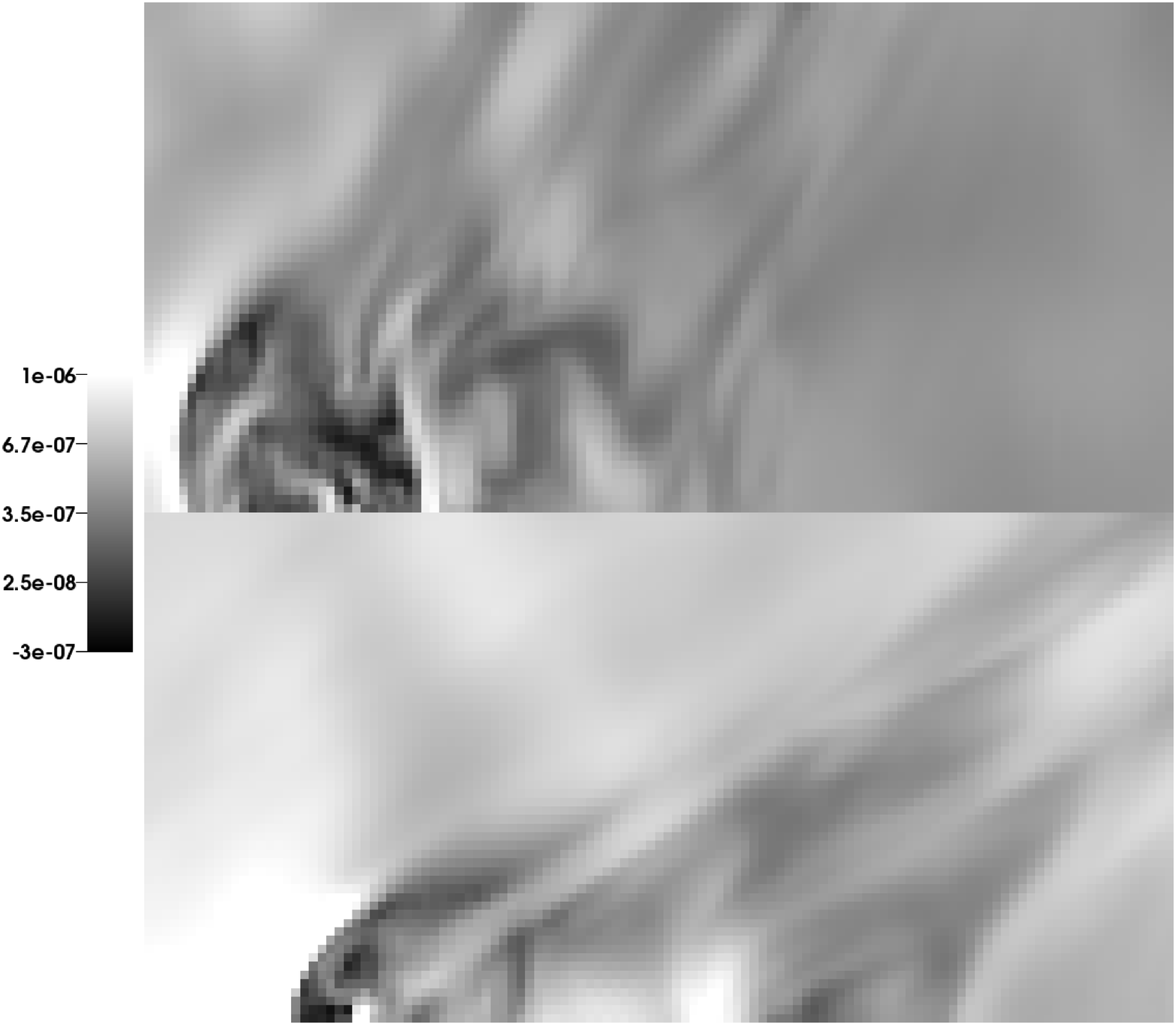}
\caption{Distribution of $B_{x}$ for Case PA 
and $B_{y}$ for Case PP at 3 Gyr in the $x-y$ plane 
centred on the galactic centre. The size of each panel is 
-50 to 190 kpc along the $x$-axis and 0 to 120 kpc along the $y$-axis. 
The ordering of the panels is:
Runs 5 ({\it top}), 5h ({\it bottom}), Cases PA ({\it left}), 
and PP ({\it right}).
The unit of the colour bar is $\sqrt{4 \pi}$ Gauss. 
Bright regions correspond to large positive values. 
}
\label{fig:low_vs_high_B}
\end{figure*}

The most significant differences between Case PA and PP are seen
in the distribution of strong magnetic field regions. 
As presented in Figure \ref{fig:B_field}, the ICM converges behind 
the galaxy and is somewhat perturbed by the ISM turbulence in Case PA. 
Since the magnetic field coupled to the ISM 
is stretched behind the galaxy due to ram pressure, 
the spatial distribution of the strongly amplified magnetic field 
is coincident with that of the stripped 
ISM. Because the fields diverge in front of the galaxy, the 
side of the galaxy exposed to 
the incoming ICM flow has weaker magnetic fields than 
those in the tail where the gas flow converges.
In Case PP, the stripped ISM forms structures flattened in the
plane of the incoming ICM flow. The tail contains regions of amplified
magnetic field. The origin of this amplification 
can be traced to the side of the galaxy exposed to the incoming ICM flow. 
It is there that the field perpendicular 
to the direction of the ram pressure is first compressed
and amplified. This layer of amplified field is very thin compared to
the size of the galaxy. The magnetic pressure in this layer 
is associated with more efficient ISM
removal in Case PP compared to Case PA. This amplified field is
subsequently advected downstream and gives rise to the pockets of
amplified magnetic field in the tail. 

\begin{figure*}
\includegraphics[scale=0.25]{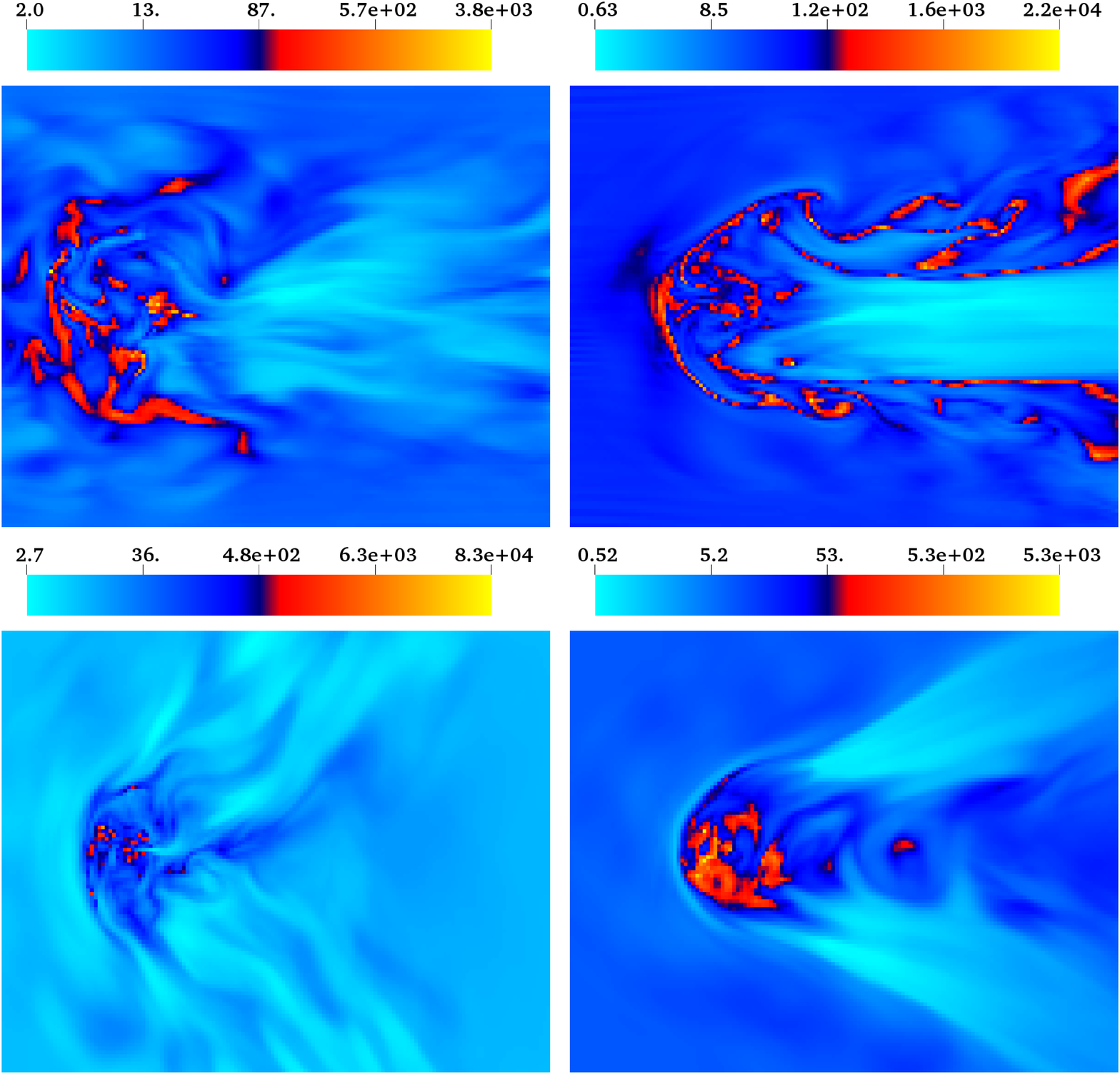}
\caption{Distribution of the plasma beta parameter in the $x-y$ plane 
centred on the galactic centre. The size of each panel is 
-80 to 220 kpc along the $x$-axis and -120 to 120 kpc along the $y$-axis. 
The snapshots are taken at 1.5 Gyr.  The ordering of the panels is:
Runs 5 ({\it left}), 5h ({\it right}), Cases PA ({\it top}), 
and PP ({\it bottom}). 
Note that the colour bar ranges vary between panels. The 
ICM flows from the left to the right.
}
\label{fig:plasma_beta}
\end{figure*}

Figure \ref{fig:B_field_profile} shows that the magnetic field in Case
PA is relatively weak ($\sim 6 \times 10^{-7}$ Gauss) on
the front side of the galaxy, i.e., for $x < -50$ kpc.  This is
consistent with Figure \ref{fig:B_field} discussed above.
This effect is due to the divergence of the field 
in the inflowing ICM. This 
field is initially aligned with the direction of the ram pressure and
begins to diverge as the ICM approaches the front of the galaxy.
The field is stronger in the tail than in the front of the galaxy, and
it is amplified to about $1.6 \times 10^{-6}$ Gauss due to the converging 
flow past the galaxy.

In Case PP, the compressed ICM and ISM is associated 
with the amplification of the
field in a thin layer in front of the galaxy. The field in this layer
is dominated by the $B_{y}$ component near $x \sim -50$ kpc 
as shown in Figure \ref{fig:B_field_profile}. 
The position of the 
maximum magnetic field strength gradually shifts
in the downstream direction as the ICM penetrates further 
into the galaxy and the ISM is removed from it.
In this case, the tail has a more strongly fluctuating magnetic field 
than the front side of the galaxy.

In Case PA, the decrease in the magnetic field strength  
around the front of the galaxy depends on the strength of turbulence 
(see Figure \ref{fig:B_field_profile}). 
Since strong turbulence in the ISM enhances the ISM mass loss, 
one may expect that the magnitude of 
the magnetic field should decrease in the front
of the galaxy (simply due to the removal of the magnetised gas from
the galaxy). However, mixing of the ICM with the ISM, combined with continuous
kinetic energy injection in the form of random motions in the galaxy, 
overcompensates for that loss of the magnetic energy inside $R_{t}$. 
For example, in Run 5, which is characterised by the strongest ISM turbulence, 
the magnitude of the magnetic field is slightly larger than in Run 0 
on the front side of the galaxy. 

In Case PP, at a given simulation time, 
the position of the strongest magnetic field 
does not vary as a function of the strength of the ISM turbulence. 
The location of the magnetic field maximum is at $x \sim -50$ kpc, 
where the ram pressure is in approximate balance
with the highly amplified magnetic and thermal pressures.
The magnetic pressure is much larger than the thermal 
and turbulent pressures, and is 
dominated by the $B_{y}$ component of the magnetic field.
Since the speed of the ICM inflow is fixed, and its kinetic energy 
is the dominant form of energy, the amplified magnetic 
field corresponds to the magnetic pressure limited by the fixed ram
pressure. The position of the peak in the magnetic field strength does not 
change significantly once the magnetic field reaches its limiting value.

Figure \ref{fig:B_field_time} shows the evolution of the magnetic field 
distribution in Run 5. This figure sheds light on the origin of the
differences between Case PA and PP shown in Figure \ref{fig:B_field_profile}.
In Case PA, as the simulation continues, the continuous loss 
of the total gas in the front of the galaxy results in 
continuous decrease of the magnetic energy density. 
This is associated with continuous divergence 
of the magnetic fields on the front side of the galaxy.
In Case PP, strong magnetic-field regions appear in the front of the
galaxy. As the galaxy loses its total gas, 
the maximum position of the strong magnetic field moves closer to the galaxy 
toward $+x$ direction.

We now comment on the evolution of the magnetic field in the tails 
shown in Figure \ref{fig:B_field_time}. 
In Case PA, the strong magnetic field in the tail 
is due to the convergence of the gas flow 
behind the galaxy. The strongest magnetic field
in the tail corresponds to the gas 
that was stripped from the galaxy in the very early stages of 
the stripping process. The region of high amplification in 
the tail is continuously receding from the galaxy as also shown 
in Figure \ref{fig:B_field}. 
Eventually, a part of the tail leaves the simulation box 
which leads to a small drop of the magnetic field magnitude 
at 6 Gyr (see the difference between the curves at 4 Gyr and 6
Gyr in Figure \ref{fig:B_field_time}). 
In Case PP, as more ISM is transported to the tail 
and the ICM flow converges behind the galaxy, 
the average strength of the magnetic field in the tail increases. However,
the collimation of the tail in Case PP is weaker than in Case PA since 
some of the stripped gas expands along the direction perpendicular to
the wind direction (i.e., in the $y$-direction). 
The magnetic fields inside $R_{t}$ 
remain relatively weak due to the continuous loss of gas 
from the galaxy. This explains the dip in the $B_{y}$ 
component inside the galaxy and relatively 
strong fields in the downstream region behind the galaxy.

\subsection{Stripping for higher ICM inflow velocity}

It is not surprising that Runs 0h and 5h lead to stronger ISM mass
loss than Runs 0 and 5. This difference is caused by nine times
stronger ram pressure in Runs 0h and 5h. 
Figure \ref{fig:ISM_high_low_comp} shows that in Runs 0h and 5h 
of both Cases PA and PP, the ISM mass 
inside $R_{t}$ drops down to 70\% of 
the original mass after about 0.5 Gyr of experiencing 
ram pressure stripping. 
This quick drop of the ISM mass is primarily driven by the
change in the outer regions of the galaxy. 

In Runs 0h and 5h for Cases PA and PP, the ram pressure 
completely dominates over magnetic fields and turbulence, 
and so the mass loss rate of the 
ISM is primarily determined by the strength of the ram pressure. 
Yet, the distributions of magnetic fields, presented in 
Figure \ref{fig:B_high_low_comp} for Runs 0h and 5h, strongly depend on 
the geometry of magnetic fields -- we observe strong differences
between Case PA and PP. This conclusion is consistent with
our previous results for lower-velocity ram pressure stripping,
i.e., for Runs 0 and 5.

The distributions of $B_{x}$ in Case PA 
and $B_{y}$ in Case PP in Runs 0h and 5h 
bear some resemblance to the corresponding quantities in Runs 0 and 5.
However, the compression of the magnetic fields 
in Runs 0h and 5h of Case PP is much stronger than in Runs 0 and 5. 
As shown in Figures \ref{fig:B_high_low_comp} and 
\ref{fig:low_vs_high_B}, 
the peak in the magnetic field on the side of the galaxy
exposed to the inflowing ICM is closer to the galaxy in Run 5h than in Run 5.

In Runs 0h and 5h of Case PA, 
the position of the lowest $B_{x}$ appears closer to the 
galactic centre than in Runs 0 and 5. 
Due to the large inflow velocity, the point where the fields begin to
diverge, and where the field strength is reduced, appears closer to
the galaxy. Consequently, at distances $-50 < x < 0$ kpc the fields are
stronger in Runs 0h and 5h than in Runs 0 and 5. 
In this case, the fast flow converging behind the galaxy generates
eddies, resulting in a locally reversed flow direction 
and negative values of $B_{x}$ (see Figure \ref{fig:low_vs_high_B}). 
Therefore, the average $B_{x}$ over $-20 < x$ kpc 
in Run 5h of Case PA is lower than in Run 5, but  
the average magnetic-field magnitude is higher in Run 5h than in Run 5.

In order to compare the dynamical impact of the amplified  
magnetic field in the runs with low and high ICM flow velocity, 
we show the distribution 
of the plasma beta in Runs 5 and 5h in Figure \ref{fig:plasma_beta}.
In Case PA, the gas in the tails is characterised by low plasma
beta. In those regions, the magnetic field 
is strong (see Figures \ref{fig:B_field} and \ref{fig:B_high_low_comp}). 
High plasma beta gas is distributed 
over the galaxy and the regions around the tail. High ram pressure
leads to a strongly converging flow behind the galaxy and plasma 
beta is much lower ($\sim$ 0.6) over a more coherent, narrower, and 
longer area in Run 5h than in Run 5.

Due to the fact that only $B_{y}$ component in the ICM is present in
Case PP, an extremely thin layer of low plasma beta appears 
on the front side of the galaxy in Run 5h 
as shown in Figure \ref{fig:plasma_beta}. 
Due to a relatively slower ICM flow,
this feature is wider in Run 5 than Run 5h. In Run 5, 
the low plasma beta ($\sim$ 2.7) 
regions are distributed over a wide area on the 
side of the galaxy exposed to the ICM flow and in the region behind
the galaxy.
In Run 5h, following the expansion of 
the gas along the $y$-axis in front of the galaxy, 
the stripped ISM and magnetic field converge behind the galaxy. 
This results in the formation of the 
low plasma beta in a wing-like structure in the $x-y$ plane 
(see Figure \ref{fig:plasma_beta}). Since high ram pressure creates 
a strongly converging flow behind the galaxy, the plasma beta is much lower 
($\sim$ 0.5) over the more coherent area behind the galaxy 
in Run 5h than in Run 5.

In general, Runs 0h and 5h have a larger fraction of volume occupied 
by the gas characterised by lower plasma beta than Runs 0 and 5.
Figure \ref{fig:plasma_beta} shows that 
plasma beta reaches $\sim ~ 0.5$ in Runs 0h and 5h and $\sim ~ 2$ 
in Runs 0 and 5.

\section{Discussion and conclusions}

We find that turbulence driven by continuous supply 
of kinetic energy to the ISM enhances the ISM mass loss in MHD simulations 
of ram pressure stripping. This effect is due to the mixing and 
efficient transport of the ISM to the outer regions of the
galaxy. This conclusion is consistent with that found in Paper I which 
discusses ram pressure stripping 
in the context of purely hydrodynamic simulations. 
However, the effects of turbulence are weaker in 
the MHD case. We find that the ISM mass loss rate depends
on the relative direction of the ICM magnetic field 
with respect to the ram pressure direction.

We show that the relative orientation of the magnetic field 
and the ram pressure direction is an important 
factor that determines the shape of the ISM stripping tails. 
This is shown in Figures 
\ref{fig:3D_para}, \ref{fig:3D_perp}, 
and \ref{fig:B_field}. When the ambient ICM has 
magnetic field components perpendicular to the ram pressure direction, 
the stripped ISM forms a wide
tail with the stripped gas distributed in the plane of the incoming
magnetic field. On the other hand, if the magnetic field is parallel
to the ram pressure direction, a long and relatively narrow tail forms only 
along that direction, and the properties of the tail are similar to those 
found in pure hydrodynamic simulations.

We also demonstrate that the initial configuration of the fields
affects the distribution of the magnetic field in the galaxy and the
ram pressure stripping tail.
When the ICM magnetic field is perpendicular to the 
ram pressure direction, the fields 
are compressed and wrapped/bent around the side of the galaxy 
exposed to the incoming ICM flow.
When the ICM magnetic field is parallel to the ram pressure direction,
the fields are weak on the side of the galaxy exposed to the wind, but  
strong magnetic fields can be found in the tail.

Radio observations of galaxies undergoing ram
pressure stripping in the Virgo cluster have been used to investigate 
magnetic fields in the ICM around late-type galaxies and in their  
ISM \citep[e.g.,][]{2007A&A...471...93W,Chyzy2008,
2010A&A...512A..36V,2010NatPh.6.520P}. 
Although there are known ellipticals experiencing ram pressure stripping 
in the Virgo cluster, such as M86 \citep{2008ApJ...688..208R} 
and NGC 4472 \citep{2011ApJ...727...41K}, there are no
reported magnetic field measurements in these cases yet.

As mentioned above, we demonstrate that 
the morphology of the tail and its magnetic field
distribution is very sensitive to the relative direction 
of the ICM magnetic field with respect to the ram pressure direction.
However, we stress that our current simulations do not imply 
that at any given time one is expected to see the whole length 
of the tails. Instead, our results show
the history of the stripping process, and the regions 
which are close to the
galaxy may best approximate instantaneous morphology of the galactic ISM and
stripping tail. However, if the magnetic field strength in these galaxies and 
in their tails could be measured, then such new observations could in
principle help us to infer 
the relative angle between the ram pressure direction and the 
direction of the locally dominant magnetic field. 
However, the interpretation of such measurements will also require a 
new set of sophisticated simulations that include tangling of the 
magnetic field and turbulence in the ICM.

Figures \ref{fig:B_field}, \ref{fig:B_field_profile}, and
\ref{fig:B_field_time} show that, when the ICM field is perpendicular
to the ram pressure direction, in the stripping tail we find thin layers 
of strong magnetic fields and a wing-like structure which is 
a filamentary region of enhanced 
magnetic fields approximately diagonal with respect to the leading 
edge of the galaxy in the $x-y$ plane. Since these
structures contain some ISM due to ICM/ISM mixing and gas stripping
(see Figure \ref{fig:B_field}), and because the ISM is 
more metal-rich than the ICM \citep{2006ApJ...639..136H,2007MNRAS.380.1554R,
2009ApJ...698..317A,2009MNRAS.399..239R,2009ApJ...696.2252J,
2011MNRAS.418.2744M}, our simulation results suggest a correlation 
between the magnetic field magnitude and the gas metallicity in the tail. 

When the ICM magnetic field is parallel to the ram pressure direction, 
the regions of strong magnetic fields also correlate with the regions 
containing high ISM fractions, which again is due to the ICM-ISM
mixing and gas stripping (see Figure \ref{fig:B_field}). 
Therefore, the high-metallicity tail, 
which could be detected using X-ray from the hot gas or optical spectroscopy
of HII regions 
\citep[e.g.,][]{1995MNRAS.277.1047R, 2008ApJ...688..931K,2010ApJ...708..946S,
2012ApJ...749...43Y,2014ApJ...786..152S}, 
should be parallel to the direction of 
the ambient ICM magnetic field. 
On the other hand, the front
side of the galaxy exposed to the ICM flow should be dominated by the ICM
(i.e., by lower metallicity gas) characterised by relatively weak 
magnetic fields because the ICM penetrates the 
galaxy more easily and the field strength is reduced as the
gas flow diverges in front of the galaxy and flows around it. 

In summary, irrespectively of the direction of the
ambient ICM fields, we expect that more strongly magnetised regions in
the tails, which are dominated by the stripped ISM, 
will have systematically higher metallicity 
than less magnetised regions in the tails. 
This conclusion should not depend on the level of 
tangling of the ambient ICM field.

Our simulations show that ram pressure stripping can not only 
lead to the metal enrichment of the ICM \citep{2008SSRv..134..363S}, 
but provide the ICM with amplified magnetic fields 
(see also \citet{2011ApJ...738...15A}). As shown in 
Figures \ref{fig:B_field} and \ref{fig:B_field_profile}, 
when the ICM magnetic fields are perpendicular to the ram pressure direction, 
parts of the stripping tail have $\sim 1.2$ times stronger magnetic field
than the ICM in Run 5. 
Similarly, when the ICM field is parallel to the ram pressure 
direction, the magnetic field in the tail is about 15\% higher than the ambient 
ICM field. These regions of the amplified magnetic field take the form of
long coherent filaments and can extend to large distances from the
galaxy (see Figure \ref{fig:B_field}). Importantly, the magnetic field
amplification increases for higher wind velocities.
These stripped ISM magnetic fields may serve as localised seed fields 
for efficient field amplification due to other mechanisms such as 
the ICM turbulence \citep[see][for a discussion]{2012ApJ...759...91C}.

We show that the dependence of the ISM mass loss rate 
on the magnetic field and turbulence strengths is insignificant 
when the ram pressure is much larger than the magnetic and turbulent
pressures (see Figure \ref{fig:ISM_high_low_comp}).
Therefore, if observations aim to explore 
the effects of the magnetic field and 
ISM turbulence on the ISM mass loss rate, 
then targets need to be cluster galaxies 
that move relatively slowly and experience 
a relatively low level of ram pressure.
However, tail morphology depends sensitively on the 
geometry of the ICM magnetic fields irrespectively how fast/slow
the galaxies are moving (see Figure \ref{fig:low_vs_high_B}).

There are multiple potentially important model components which could 
be investigated in future numerical simulations of ram pressure 
stripping.
For example, our simulations only consider uniform fields in the ICM, 
while the realistic ICM is turbulent and the ICM fields are
not uniform \citep{2005MNRAS.358..139F,schekochihin:056501,
2006MNRAS.366.1437S,2008MNRAS.384.1567M,2010ApJ...721.1105B}. 
Moreover, the density and velocity fluctuations in 
the ICM flow may also further complicate the stripping process. 
The ram pressure is proportional to the density and 
the square of galactic velocity with respect to the local ICM. 
Therefore, the stripping efficiency, 
tail morphology, and its magnetic fields may be affected by the ICM
density and velocity fluctuations. 
The ICM turbulence can also alter the mixing 
between the stripped ISM and the ICM \citep{annurev-fluid-010313-141357}.

We do not include realistic modelling of turbulence in the
ISM. For example, in realistic systems, turbulence will be driven in
part by supernovae and active galactic nuclei (AGN) \citep{2003ARA&A..41..191M,
2004ARA&A..42..211E,2013MNRAS.430.1516H}, which we do not explicitly include.
Moreover, the interplay between turbulence and ram pressure stripping 
is further complicated by the possibility that 
ram pressure stripping itself could partially suppress 
star formation and AGN activities 
\citep[e.g., ][]{1992MNRAS.255..346B,1999MNRAS.309..161M,
2008A&A481..337K,2012ApJ...745...13S}. Such complexities are beyond
the scope of the current paper.

Finally, in addition to the refinement methodology, 
other processes such as radiative cooling, 
heat conduction, and self-gravity 
of the gas can alter the mass loss rates, 
star formation in the stripping tail, 
and the morphology of the gas and magnetic field distributions 
\citep[e.g.][]{2004ApJ...606L.105A,2010ApJ...717..147S,2010ApJ...722..412Y,
2012MNRAS.422.1609T,2012A&A...545A.142B}. 
In particular, heat conduction between the stripped 
warm ISM and the hot ICM might be an important process
that shapes the tail morphology 
\citep{MNRAS238.1247, 1990ApJ...358..375B,2012ApJ.748.24L}. 
Since heat conduction is anisotropic in magnetised plasmas, 
including heat conduction can also affect cooling in the MHD case 
\citep[e.g.,][]{2004JKAS...37..575A,2008ApJ...678..274O,2012ApJ.748.24L}.

\section*{Acknowledgements}

We are grateful to Karen Yang and Dongwook Lee for useful
discussions. We appreciate the input of an anonymous referee.
MR acknowledges NSF grant AST 1008454 and NASA ATP 12-ATP12-0017. 
The software used in this work was in part developed by the DOE
NNSA-ASC OASCR Flash Center at the University of Chicago.  
This work used the Extreme Science and Engineering Discovery
Environment (XSEDE), which is supported by National Science Foundation  
grant number OCI-1053575.

\appendix

\section{Initial magnetic fields in the weak ram pressure stripping case}

We present results for different strengths of the initial magnetic
fields in the computational domain and at the inflow boundary. The
fields are initially constant in the whole computational domain and
constant throughout the simulation at the inflow boundary.
The evolution of the fields that enter the simulation box in Cases PA
and PP is different in these two cases 
in part because the field strength applied at the
boundary transforms differently depending on the 
relative angle of the magnetic field with respect to the inflow direction.

The impact of different field orientations on the field entering the
simulation domain is shown in 
Figures \ref{fig:B_field_profile} and \ref{fig:B_field_time}. 
In particular, Figure \ref{fig:B_field_time} shows that the magnitude of the 
magnetic field entering the simulation box is about $1.44 \times 10^{-6}$
Gauss in Case PA and about $2.04 \times 10^{-6}$ Gauss in Case
PP. This difference occurs despite the fact that, in both cases, we assume
that the initial ISM/ICM magnetic field strengths and the strength of the
field at the boundary are all identical (i.e., $1.44 \times 10^{-6}$ Gauss).

In order to eliminate these differences in the level of the field
entering the computational box, we perform Case PA runs with a stronger
magnetic field and use $2.04 \times 10^{-6}$ Gauss instead of 
the original magnitude of $1.44 \times 10^{-6}$ Gauss. As in the
original runs, we initialise the same magnetic field in
both the ISM and ICM (this time $2.04 \times 10^{-6}$ Gauss),
keeping all other parameters unchanged.
In these new Case PA runs, the effective strength of the field 
that initially enters the computational domain 
approximately matches that in the original run in
Case PP. Thus, these new runs allow us 
to isolate the effects of the geometry of the initial/inflowing fields 
on the amount of gas stripping and the strength of the resulting 
magnetic field.

\begin{figure}
\includegraphics[scale=0.43]{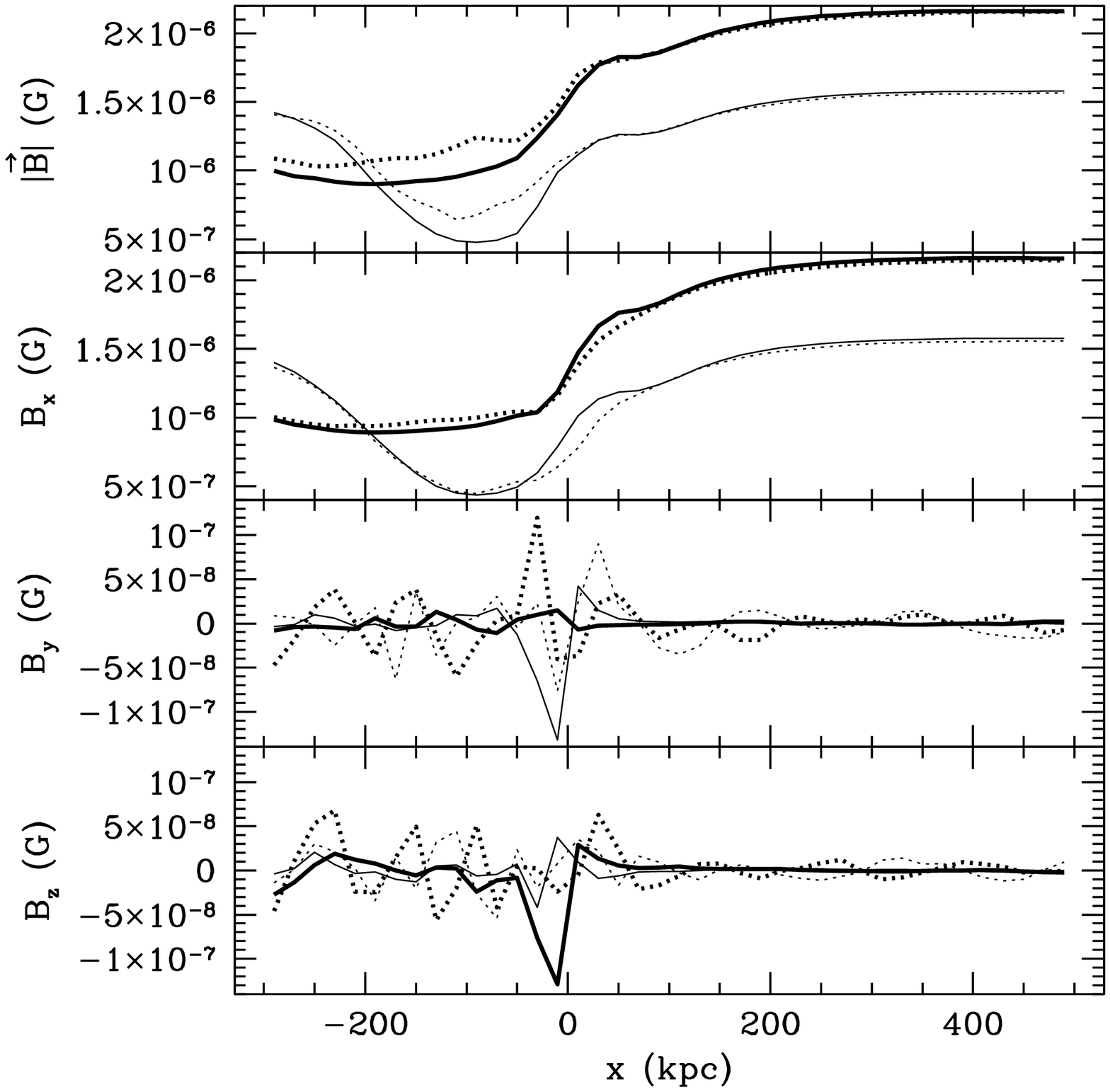}
\caption{
Mass-weighted average magnetic field magnitude,
$x$, $y$, and $z$-components (from top to bottom) along $x$-axis 
for Case PA. The magnitudes are measured inside a cylinder 
of radius 120 kpc in radius at 5 Gyr. The original runs are 
shown as {\it thin lines} and the new runs as {\it thick lines}. 
The results are presented for 
Runs 0 ({\it solid line}) and 5 ({\it dotted line}). 
The bin size along $x$-axis is 20 kpc.
}
\label{fig:B_field_profile_appendix1}
\end{figure}

Figure \ref{fig:B_field_profile_appendix1} shows the distribution of 
the evolved field in the new runs. Although the general trends seen in 
the new runs are not different from those found 
in the original runs, the new runs demonstrate that 
the stronger initial and boundary magnetic fields lead to generally stronger
magnetic fields 
(including in the regions directly in front of the galaxy). However, as the
dynamical effects of the magnetic field are now stronger than in the original
runs, the buildup of the field just ahead of the galaxy affects the
field at a larger upstream distance away from the galaxy. 
This results in lower fields in that region compared to the original run. 

\begin{figure}
\includegraphics[scale=0.43]{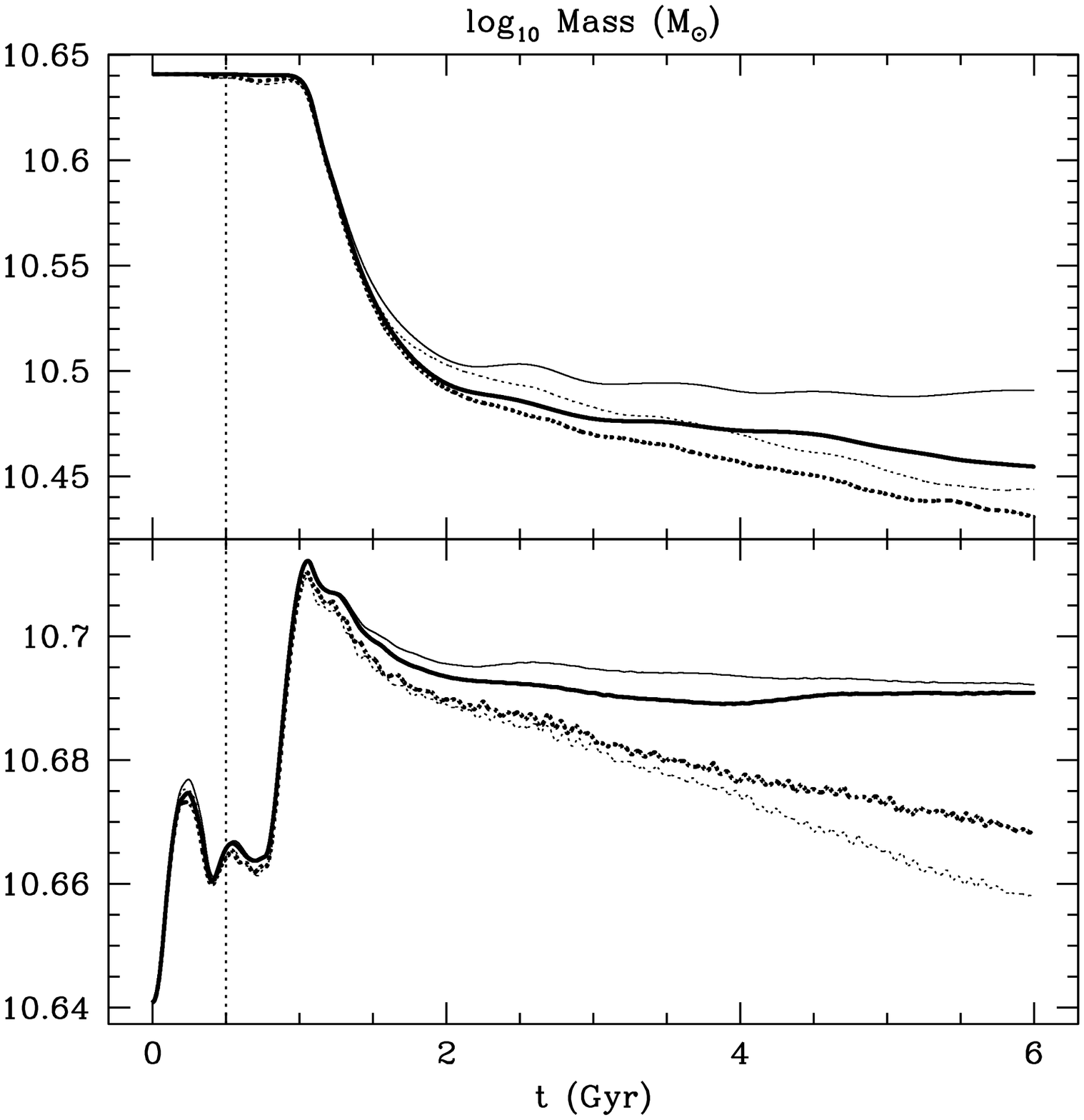}
\caption{Mass evolution of the intrinsic ISM mass 
({\it top}) and the total gas mass ({\it bottom}) 
inside $R_{t}$ in Case PA. With the exception of the vertical dotted 
line that corresponds to 0.5 Gyr when the ICM begins to flow into
simulation box, all line styles have 
the same meaning as in \ref{fig:B_field_profile_appendix1}.}
\label{fig:mass_appendix1}
\end{figure}

Figure \ref{fig:mass_appendix1} shows that the stronger 
field in the new runs results in more efficient removal of the 
ISM beyond $R_{t}$ compared to that seen in the original runs.
The figure also compares the evolution of the total 
gas mass in the galaxy between the new and old runs. While the differences 
between the values of the total mass inside $R_{t}$ are 
not significant in Run 0, the stripping is less efficient for stronger
fields in Run 5 (i.e., when turbulence is stronger). However, 
the difference between Run 0 and 5 is smaller in the new runs than in 
the original runs because the impact of different turbulence strength  
is overwhelmed by the effect of the magnetic field (which is stronger
in the new runs).


\begin{figure}
\includegraphics[scale=0.43]{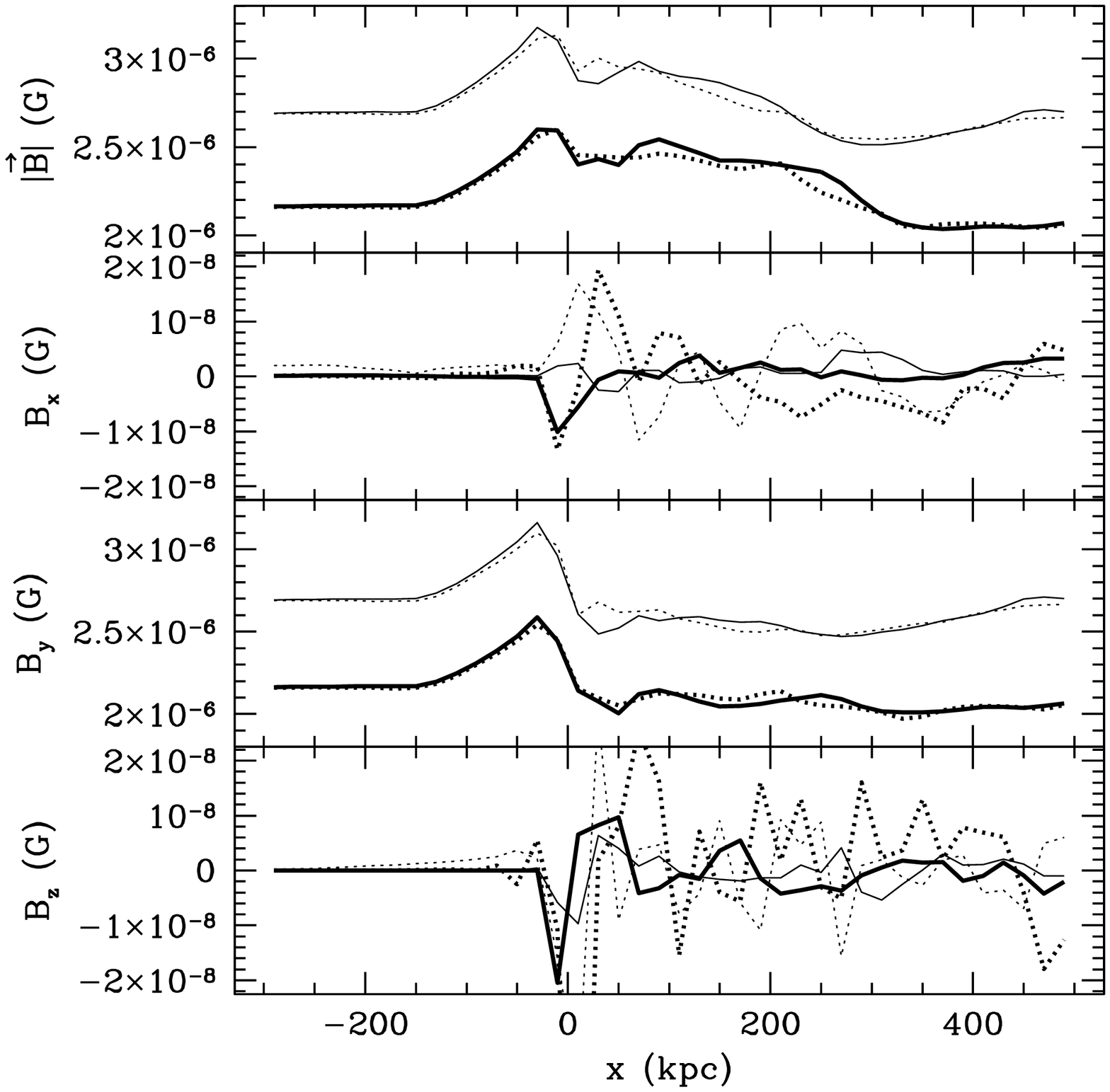}
\caption{
Mass-weighted average magnetic field magnitude,
$x$, $y$, and $z$-components (from top to bottom) 
along $x$-axis for
Case PP. The magnitudes are measured inside a cylinder 
of radius 120 kpc in radius at 5 Gyr. 
The original runs are shown as {\it thin lines}
and the new runs for the strong ram pressure case 
as {\it thick lines}. The results are presented for 
Runs 0h ({\it solid line}) and 5h ({\it dotted line}). 
The bin size along $x$-axis is 20 kpc.}
\label{fig:B_field_profile_appendix2}
\end{figure}
\begin{figure}
\includegraphics[scale=0.43]{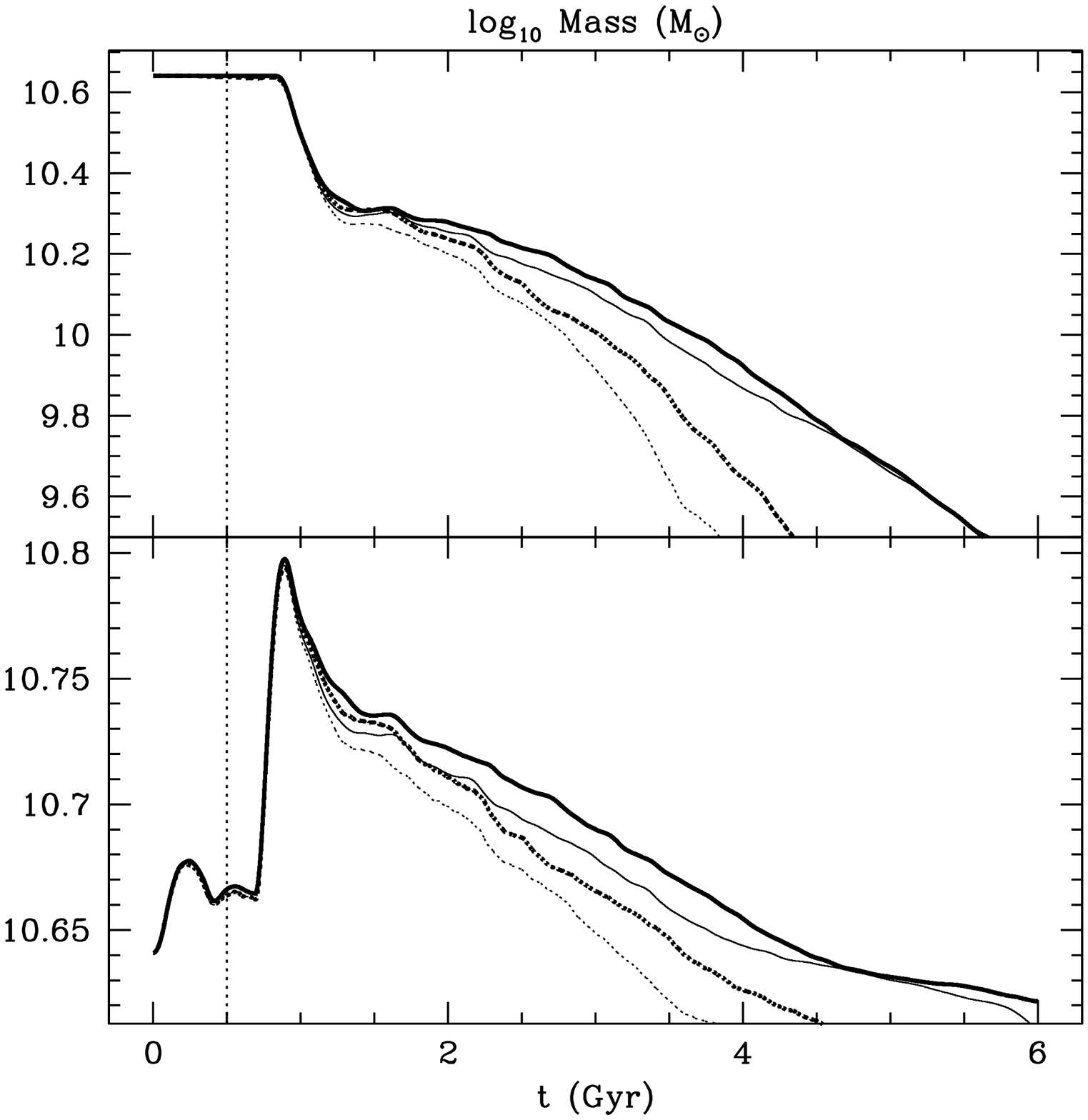}
\caption{
Mass evolution of the intrinsic ISM mass ({\it top}) and the total gas mass 
({\it bottom}) inside $R_{t}$ for the strong ram pressure in Case PP.
With the exception of the vertical dotted 
line that corresponds to 0.5 Gyr when the ICM begins to flow into
simulation box, all line styles have the same meaning 
as in \ref{fig:B_field_profile_appendix2}.}
\label{fig:mass_appendix2}
\end{figure}

The new runs also allow us to check if the large loss of the ISM 
found in Case PP is solely caused by the differences between the
magnetic field strengths in the inflowing ICM in Cases 
PA and PP. For example, the mass of the ISM retained inside $R_{t}$ 
in the new Run 5 of Case PA is about ${\rm 10^{10.43} M_{\odot}}$ 
at 6 Gyr, which is about 7\% more than in Case PP 
(see Figure A2 and Figure 8). 
This difference is quite similar to the one between Case PP and 
PA in the original runs. 
Since the strength of the inflowing fields (and the initial ones) 
in the original Case PP and the new Case
PA is now very similar, the differences in the ISM stripping rates 
can now be attributed to the geometry of the magnetic field 
rather than the field strengths. 
Similar conclusions can be drawn with regard to the total gas 
stripping. Specifically, Figure \ref{fig:mass_appendix1} 
shows that the removal of the total gas from within $r<R_{t}$ 
is somewhat more efficiently suppressed 
due to deeper penetration of the ICM into the galaxy 
in the new Case PA compared to the original Case PP (see Figure
\ref{fig:mass_gas}).
Therefore, the differences between the original Case PP 
and the new run of Case PA are even larger 
than between Case PP and Case PA in the original runs, 
demonstrating that these differences in the total mass stripping 
rates can be attributed to the
geometry of the field rather than its strength.

\section{Initial magnetic fields in the strong ram pressure stripping case}

Figure \ref{fig:B_high_low_comp} shows that the overall strength of
the magnetic fields in Runs 0h and 5h of Case PP is 
larger than in Runs 0 and 5.
The initial field strength in the ISM/ICM, as well as the strength 
of the field imposed at the inflow boundary, 
is the same in all of these runs. 
The difference in the strength of the evolved field 
is due to the fact that: (a) the field imposed 
at the boundary differs from that 
actually entering the computational domain (in the high ram pressure
case the latter is slightly higher), and (b) the dynamical interaction
of the inflowing ICM with the galaxy is different 
due to the different levels of ram pressure. 
In order to disentangle these two possible effects, we perform extra runs 
of strong ram pressure for Case PP in which we lower the initial field 
(and the field imposed at the boundary) to 
$1.15 \times 10^{-6}$ Gauss. 
All other parameters are unchanged in these new runs. 
Upon entering the computational box, the field strength now
approximately matches that in the original lower ram pressure run of Case PP.

We find that the general tendency to form a strong magnetic field layer
in front of the galaxy is still present in the new runs.
Figure \ref{fig:B_field_profile_appendix2} shows 
that a strong magnetic field layer forms around x $\sim -30$ 
kpc as in the original high ram pressure run. 
However, the strong ram pressure 
pushes the gas with amplified magnetic field deeper into the galaxy than in
the weak ram pressure stripping case, 
resulting in the amplification of the magnetic field over a wider
range of distances. This conclusion agrees with 
our previous conclusion for the original 
high ram pressure run (see Figure \ref{fig:B_high_low_comp}). 

Figure \ref{fig:mass_appendix2} presents the evolution 
of the intrinsic ISM mass and the total gas 
mass inside $R_{t}$ in the new runs and original runs of Case PP. 
Decreasing the initial magnetic field strength 
reduces ability of the incoming ICM flow to strip the ISM 
in both Run 0h and 5h. This tendency is more pronounced in 
Run 5h than in Run 0h. This attenuated ISM loss is also consistent with 
the larger total gas mass in the galaxy.

The comparison of Figure \ref{fig:mass_appendix2} and
Figure \ref{fig:ISM_high_low_comp} confirms that 
the larger ISM mass loss observed in the strong ram pressure stripping case 
(i.e., Runs 0h and 5h) is indeed mainly determined 
by the velocity of the ICM inflow rather than
the differences due to the dynamical effect of the magnetic field
(i.e., rather than 
the different strength of magnetic fields in the inflowing ICM). 
Even after lowering the initial and boundary magnetic field strengths to 
ensure that the actual field entering the computational domain matches 
the field strength in the weak ram pressure stripping case, 
we find that the ISM loss in the strong ram pressure stripping 
case is far above that found in the weak ram pressure stripping case.

\end{document}